\documentclass[%
reprint,
]{revtex4-1}

\usepackage{graphicx}
\usepackage{physics}
\usepackage{siunitx}
\usepackage{dcolumn}
\usepackage{longtable}
\usepackage{makecell}
\usepackage{multirow}

\begin{document}

\title{Momentum-Space Unitary Coupled Cluster and Translational Quantum Subspace Expansion for Periodic Systems on Quantum Computers}

\author{David Zsolt Manrique}
\email{david.zsolt.manrique@cambridgequantum.com}
\author{Irfan T. Khan}
\author{Kentaro Yamamoto}
\author{Vijja Wichitwechkarn}
\author{David Mu\~{n}oz Ramo}
\email{david.munoz.ramo@cambridgequantum.com}

\affiliation{Cambridge Quantum Computing Ltd.\\ 9a Bridge Street, CB2 1UB Cambridge\\ United Kingdom}

\begin{abstract}
We demonstrate the use of the Variational Quantum Eigensolver (VQE) to simulate solid state crystalline materials. We adapt the Unitary Coupled Cluster ansatz to periodic boundary conditions in real space and momentum space representations and directly map complex cluster operators to a quantum circuit ansatz to take advantage of the reduced number of excitation operators and Hamiltonian terms due to momentum conservation. 
To further reduce required quantum resources, such as the number of UCCSD amplitudes, circuit depth, required number of qubits and number of measurement circuits, we investigate a translational Quantum Subspace Expansion method (TransQSE) for the localized representation of the periodic Hamiltonian. Additionally, we also demonstrate an extension of the point group symmetry based qubit tapering method to periodic systems. We compare accuracy and computational costs for a range of geometries for 1D chains of dimerized hydrogen, helium and lithium hydride with increasing number of momentum space grid points and also demonstrate VQE calculations for 2D and 3D hydrogen and helium lattices. 
Our presented strategies enable the use of near-term quantum hardware to perform solid state simulation with variational quantum algorithms. 
\end{abstract}


\maketitle

\section{Introduction}

The simulation of molecules and materials has been identified for a long time as one of the key areas that will benefit from quantum computing. 
This is due to the natural way in which electron correlations may be implemented on quantum hardware, with an exponential reduction in the memory required to express the correlated wavefunction. 
As quantum circuits for calculations of molecular systems have comparatively lower requirements than for other applications, they are well suited for implementation on current and near-term quantum hardware (the so-called NISQ machines), which is characterized by low numbers of qubits and high levels of operational noise. 

Variational approaches like the Variational Quantum Eigensolver (VQE) \cite{peruzzo} have become the standard for use on NISQ hardware. 
Most of the theoretical developments have focused on the calculation of energies for molecular systems, while the research on solid state systems has mostly been taking place on the use of model Hamiltonians like the Hubbard model or approaches using periodic boundary conditions (PBC) but requiring fault tolerant machines \cite{google_pbc}. 
However, implementations of variational quantum algorithms for PBC systems have been scarce; 
we acknowledge band structure calculations for simplified tight-binding Hamiltonians \cite{cerasoli_silicon}, or total energy calculations for hydrogen chains \cite{liu_pbc}. 
PBC is one of the main workhorses in the calculation of the electronic structure and properties of extended systems, including crystals, large polymers, amorphous phases, surfaces and interfaces, via its combination with Density Functional Theory in its many flavours. 
Therefore, quantum algorithms for solid state calculations that integrate this approximation are of great interest for their possibility of enabling the efficient use of wavefunction methods in periodic models, thus providing a competitive alternative to the different density functionals currently in use. 
This paper aims to contribute to this research area by presenting an adaptation of the VQE algorithm with explicit calculation of the correlation energy to different kinds of periodic systems including 1D, 2D and 3D lattices, as well as a new variational algorithm based on the Quantum Subspace Expansion (QSE) \cite{linear1, linear2} and Non-orthogonal VQE \cite{Huggins_2020}.

Extension of the VQE algorithm to PBC settings requires modifications at many different parts of the machinery of this method, as a result of the need to include the dependence with respect to the reciprocal space associated to the system being calculated. 
The qubit encoding has to be designed in such a way that each qubit is associated not only to a spin-orbital, but also to a particular k-point from the reciprocal space sampling grid considered. 
In addition, the electronic integrals required to define the second-quantized Hamiltonian have now k-point dependence, as well as the ansatz used to construct the correlated wavefunction.

Our construction of a PBC-adapted VQE algorithm builds from techniques already investigated in traditional quantum chemistry and materials science in the framework of wavefunction methods like configuration interaction and coupled cluster theory applied to periodic systems \cite{bartlett_pbc, chan_ucc_pbc, alavi_fci_pbc, gruneis_cc_pbc1, gruneis_cc_pbc2}. 
These techniques have been investigated with the purpose of approaching electron correlation in a controlled way, avoiding the uncertainty associated to the use of Density Functional Theory approaches. 
This issue is particularly important in the study of materials with strong electron correlations; representative examples are Mott-Hubbard insulators or molecular crystals with strong van der Waals interactions. 
However, development and adoption of these methods by the materials modeling community has been slow due to their very high computational cost in comparison with the requirements for molecular system calculations. 
Quantum computing may provide a way to make these approaches feasible to the wider community, once advances in the capabilities of quantum hardware are mature enough. 
It may also enable the development of novel ideas in the materials modeling community in the framework of the study of the performance of the Unitary Coupled Cluster Singles and Doubles (UCCSD) ansatz in a periodic setting. Benchmark studies on classical machines suggest that UCCSD performs well at the description of many molecular multireference states compared with the typical non-unitary ansatz used in calculations on classical machines \cite{knowles_ucc}; its behaviour in a periodic multireference system remains an open question. In this sense, our work provides a starting point to address these complex problems.

In the following sections, we provide a description of the method, starting by our approach to encode the periodic wavefunction in a qubit register. 
This is followed by a summary of the method for constructing a PBC-adapted ansatz based on the UCCSD operator. 
Next, we explain our work to simplify the state preparation procedure by means of the Translational Quantum Subspace Expansion approach, a key subject given the extra complexity of a periodic wavefunction in comparison to the standard one. 
We wrap up discussion by showing test calculations on several simple systems, analyzing the resources needed for periodic calculations on quantum computers, and laying out future challenges ahead.

\section{Method}

 Solid state simulations are often performed with plane wave basis sets, in particular mean-field simulations, for which the necessary integrals can be computed efficiently with the Fast Fourier Transform method or other grid based methods. 
 To achieve accurate results for systems with localized features the number of plane wave basis functions needs to be relatively large. This problem is often mitigated by replacing the orbitals and electrons present in the atomic cores with ultrasoft pseudopotentials or other post-processing techniques involving basis set rotations \cite{gruneis_pbc_ano}. 
 To perform solid state simulations with periodic boundaries alternative approaches have been developed using localized basis functions, such as Gaussian basis sets, which have been extensively used for simulations in areas like heterogeneous catalysis, functional materials engineering, 2D structures, electron transport or high-pressure phases, among others \cite{delgado_cry17, warner_cry17, valov_cry17, constantinescu_cry17, islam_cry17, rohlfing_cry17, nagarajan_cry17, zhu_cry17, polaron_cry17, lunghi_cry17, raty_cry17, pinotsi_cp2k, semino_cp2k, imoto_cp2k, chan_pyscf_example, wang_pyscf_example}.  
 The computation of integrals with localized basis functions is more expensive than using a plane wave basis set, but the number of basis functions is significantly smaller, which is favorable for quantum computations in the NISQ era. Similarly to the molecular VQE methodology, the VQE for systems with periodic boundary conditions requires a set of electronic integrals and orbitals. 
 In this work we employ localized basis functions which are constructed from Gaussian functions and adapted to translational symmetry. 
 The required periodic integral evaluations, the Hartree--Fock and higher level reference calculations, such as the classical k-point dependent Coupled Cluster Singles and Doubles (CCSD) and the Full Configuration Interaction method (FCI) were performed by the classical package PySCF 1.7 \cite{pyscf}. 
 There are other available codes that can provide integrals and perform classical simulations for periodic systems \cite{crystal17, cp2k, gaussian16} and the detailed methodology is described in Refs. \cite{chan_ucc_pbc, evarestov2012hartree}. 
 
 Two typical representations of the electronic Hamiltonian are based on expansions with real space localized orbitals and on expansions with orbitals in momentum or reciprocal space. 
 For a crystal structure that is characterized by its primitive cell and the corresponding lattice vectors $\mathbf{a}_1$, $\mathbf{a}_2$ and $\mathbf{a}_3$, the second-quantized electronic Hamiltonian adopts the following form:
\begin{multline*}
\hat{H}_R = 
\sum_{\mathbf{R}_p\mathbf{R}_q}
\sum_{pq}
\sum_{\sigma} 
h^{\mathbf{R}_pp}_{\mathbf{R}_qq}  
\hat{a}^{\dagger}_{\mathbf{R}_{p}p\sigma}\hat{a}_{\mathbf{R}_{q}q\sigma} 
+ \\ + 
\frac{1}{2}
\sum_{\mathbf{R}_{p}\mathbf{R}_{q}\mathbf{R}_{r}\mathbf{R}_{s}}
\sum_{\substack{pqrs, \\ \sigma\sigma'}}
g^{\mathbf{R}_{p}p,\mathbf{R}_{r}r}_{\mathbf{R}_{q}q,\mathbf{R}_{s}s} \hat{a}_{\mathbf{R}_{p}p\sigma}^{\dagger}\hat{a}_{\mathbf{R}_{r}r\sigma'}^{\dagger}\hat{a}_{\mathbf{R}_{s}s\sigma'}\hat{a}_{\mathbf{R}_{q}q\sigma}
\end{multline*}

where $\mathbf{R}_p = l_p^{(1)} \mathbf{a}_1 + l_p^{(2)} \mathbf{a}_2 + l_p^{(3)} \mathbf{a}_3 $ is a lattice vector that labels translated primitive cells from the origin, $l_p^{(1)}$,$l_p^{(2)}$ and $l_p^{(3)}$ are integers, $\hat{a}_{\mathbf{R}_{p}p\sigma}$, $\hat{a}^{\dagger}_{\mathbf{R}_{p}p\sigma}$ are annihilation and creation fermion operators for an orbital $p$ and spin $\sigma$ in the primitive cell at $\mathbf{R}_{p}$, and $h^{\mathbf{R}_pp}_{\mathbf{R}_qq}$ and $g^{\mathbf{R}_{p}p,\mathbf{R}_{r}r}_{\mathbf{R}_{q}q,\mathbf{R}_{s}s}$ are the one- and two-electron integrals \cite{evarestov2012hartree, pyscf, chan_ucc_pbc}. 
If the localized orbitals are orthogonal, such as the Wannier orbitals, then the ladder operators satisfy the anti-commutation relations. 
In this general form the number of terms of the Hamiltonian in a supercell that contains $L_1,L_2,L_3$ primitive cells in the three respective dimensions scales with $O(L^4N^4)$ where $L=L_1 L_2 L_3$ and $N$ is the number of orbitals within a primitive cell. 
The relation between the real space lattice and the reciprocal lattice representations is given by the discrete Fourier expansion \cite{evarestov2012hartree}
\begin{equation}
\hat{a}_{\mathbf{R}p\sigma}=\frac{1}{\sqrt{L}}\sum_{\mathbf{k}}e^{i\mathbf{k}\cdot\mathbf{R}}\hat{c}_{\mathbf{k}p\sigma}
\label{eq:expansion}
\end{equation}
where $\hat{c}_{\mathbf{k}p\sigma}$ is an annihilation operator for the orbital $p$, spin $\sigma$ and label $\mathbf{k}=\frac{k^{(1)}}{L_1} \mathbf{b}_1 + \frac{k^{(2)}}{L_2} \mathbf{b}_2 + \frac{k^{(3)}}{L_3} \mathbf{b}_3$, $\mathbf{b}_1$,$\mathbf{b}_2$ and $\mathbf{b}_3$ are the reciprocal lattice vectors and $k^{(1)}$, $k^{(2)}$ and $k^{(3)}$ are integers such that $-\frac{L_{\alpha}}{2}<k^{(\alpha)}\leq\frac{L_{\alpha}}{2}$ for $\alpha=1,2,3$. 
After the transformation the Hamiltonian takes the form of
\begin{multline*}
\hat{H}_K = 
\sum_{\mathbf{k}} 
\sum_{pq} 
\sum_{\sigma} 
h^{\mathbf{k}p}_{\mathbf{k}q}  
\hat{c}^{\dagger}_{\mathbf{k}p\sigma}\hat{c}_{\mathbf{k}q\sigma} 
+ \\ +
\frac{1}{2}
\sum'_{\mathbf{k}_{p}\mathbf{k}_{q}\mathbf{k}_{r}\mathbf{k}_{s}}
\sum_{\substack{pqrs, \\ \sigma\sigma'}}
g^{\mathbf{k}_{p}p, \mathbf{k}_{r}r}_{\mathbf{k}_{q}q, \mathbf{k}_{s}s} \hat{c}_{\mathbf{k}_{p}p\sigma}^{\dagger}\hat{c}_{\mathbf{k}_{r}r\sigma'}^{\dagger}\hat{c}_{\mathbf{k}_{s}s\sigma'}\hat{c}_{\mathbf{k}_{q}q\sigma}
\end{multline*}
where $h^{\mathbf{k}p}_{\mathbf{k}q}$ and $g^{\mathbf{k}_{p}p, \mathbf{k}_{r}r}_{\mathbf{k}_{q}q, \mathbf{k}_{s}s}$ are the transformed one- and two-electron integrals of the periodic system. 
Due to the conservation of crystal momentum, after the transformation the one electron integrals become diagonal in $\mathbf{k}$ and the non-zero two-electron integrals obey $\mathbf{k}_p + \mathbf{k}_r - \mathbf{k}_q - \mathbf{k}_s = \mathbf{G}$ where $\mathbf{G}$ is a reciprocal lattice vector \cite{chan_ucc_pbc}. The prime symbol in the sum indicate that it is expanded according to this rule. 
Consequently, a large amount of 2-electron operators from the Hamiltonian are discarded and as a result the number of terms in $\hat{H}_K$ reduces to $O(L^3N^4)$. 

The mean-field calculations are performed in momentum space as the one electron density matrix as well as the Fock matrix are also diagonal in $\mathbf{k}$, therefore PySCF's periodic Hartree--Fock calculation directly provides the crystal orbitals, which are the periodic analogues of the molecular orbitals \cite{chan_ucc_pbc}. In addition, PySCF also provides the corresponding occupations, orbital energies $\varepsilon_{p}(\mathbf{k})$ and the integrals $h^{\mathbf{k}p}_{\mathbf{k}q}$, $g^{\mathbf{k}_{p}p, \mathbf{k}_{r}r}_{\mathbf{k}_{q}q, \mathbf{k}_{s}s}$ directly in the momentum space. 

In the following subsections, we provide details about the qubit encoding scheme, the design of the UCCSD-PBC ansatz with translational symmetry and our derivation of the TransQSE method.

\subsection{Qubit encoding of quantum numbers}

\begingroup
\squeezetable
\begin{table}
    \caption{Qubit encoding for the Hydrogen chain $d=\SI{0.75}{\angstrom}$, $L=2$ case in momentum space and $a=2.5d$.}
    \label{tab:encodings_H2k2}
    \begin{ruledtabular}
        \begin{tabular}{c|ccccccc|r}
        $q^{(K)}$ & $\mathbf{k}$ & $k^{(1)}$ & $k^{(2)}$ & $k^{(3)}$ & orbital, $p$ & spin & occ.& $\varepsilon_{p}(\mathbf{k})$\footnote{Orbital energies} (Ha)\tabularnewline
        \hline 
        $0$ & $\frac{\pi}{a}\left(0,0,0\right)$ & $0$ & $0$ & $0$ & $0$ & $\uparrow$ & $1$ & \multirow{2}{*}{$-0.7159337$}\tabularnewline
        \cline{1-8} 
        $1$ & $\frac{\pi}{a}\left(0,0,0\right)$ & $0$ & $0$ & $0$ & $0$ & $\downarrow$ & $1$ & \tabularnewline
        \hline 
        $2$ & $\frac{\pi}{a}\left(0,0,0\right)$ & $0$ & $0$ & $0$ & $1$ & $\uparrow$ & $0$ & \multirow{2}{*}{$1.5283639$}\tabularnewline
        \cline{1-8}
        $3$ & $\frac{\pi}{a}\left(0,0,0\right)$ & $0$ & $0$ & $0$ & $1$ & $\downarrow$ & $0$ & \tabularnewline
        \hline 
        $4$ & $\frac{\pi}{a}\left(1,0,0\right)$ & $1$ & $0$ & $0$ & $0$ & $\uparrow$ & $1$ & \multirow{2}{*}{$-0.392922$}\tabularnewline
        \cline{1-8} 
        $5$ & $\frac{\pi}{a}\left(1,0,0\right)$ & $1$ & $0$ & $0$ & $0$ & $\downarrow$ & $1$ & \tabularnewline
        \hline 
        $6$ & $\frac{\pi}{a}\left(1,0,0\right)$ & $1$ & $0$ & $0$ & $1$ & $\uparrow$ & $0$ & \multirow{2}{*}{$0.3863021$}\tabularnewline
        \cline{1-8} 
        $7$ & $\frac{\pi}{a}\left(1,0,0\right)$ & $1$ & $0$ & $0$ & $1$ & $\downarrow$ & $0$ & \tabularnewline
        \end{tabular}
    \end{ruledtabular}
\end{table}
\endgroup

In a molecular VQE calculation the qubit registers are associated with the spin-orbitals of the molecule. With the Jordan--Wigner encoding each qubit encodes the electron occupation of the corresponding spin-orbital. This correspondence remains the same for the periodic systems as well, but the spin-orbitals are adapted to the translational symmetry. 
In particular, in the momentum space the so-called crystal spin-orbitals have the additional label $\mathbf{k}$, and similarly in the real space the spin-orbitals have the extra label $\mathbf{R}$.  
Consequently in both cases the number of qubits required to represent the periodic wavefunction in a supercell with size $L$ is $2N \times L$. 
Similarly, the number of electrons is $N_e\times L$ where $N_e$ is the number of electrons in the primitive cell. 
The mapping between the qubit index and spin-orbitals in the real space representation, for a supercell with the size $L_1,L_2,L_3$ is 
\begin{equation}
q_R : (\mathbf{R},p,\sigma) \rightarrow 2 N \left( l^{(1)} + L_1 l^{(2)} + L_1 L_2 l^{(3)} \right ) + 2p + \sigma,    
\end{equation}
and the mapping in momentum space is
\begin{equation}
q_K : (\mathbf{k},p,\sigma) \rightarrow 2 N \left( \tilde{k}^{(1)} + L_1 \tilde{k}^{(2)} + L_1 L_2 \tilde{k}^{(3)} \right)  + 2p + \sigma
\end{equation}
where $\tilde{k}^{(\alpha)}=k^{(\alpha)} \mod L_{\alpha}$ and we are using the convention that the orbital index $p$ starts from $0$, and $\sigma = 0$ represents the up and $\sigma = 1$ represents the down-spin. 
Table \ref{tab:encodings_H2k2} shows an example of qubit encoding for a hydrogen chain with two atoms per primitive cell, the STO-3G basis set (2 spin-orbitals per H atom) and two k-points in the reciprocal space grid. 
\begingroup
\squeezetable
\begin{table}
    \caption{Qubit encoding for the Hydrogen chain $d=\SI{0.75}{\angstrom}$, $L=3$ case in momentum space and $a=2.5d$.}
    \label{tab:encodings_H2k3}
    \begin{ruledtabular}
        \begin{tabular}{c|ccccccc|r}
        $q^{(K)}$ & $\mathbf{k}$ & $k^{(1)}$ & $k^{(2)}$ & $k^{(3)}$ & orbital, $p$ & spin & occ. & $\varepsilon_{p}(\mathbf{k})$\footnote{Orbital energies} (Ha)\tabularnewline
        \hline 
        $0$ & $\frac{2\pi}{3a}\left(0,0,0\right)$ & $0$ & $0$ & $0$ & $0$ & $\uparrow$ & $1$ & \multirow{2}{*}{$-0.7568523$}\tabularnewline
        \cline{1-8} 
        $1$ & $\frac{2\pi}{3a}\left(0,0,0\right)$ & $0$ & $0$ & $0$ & $0$ & $\downarrow$ & $1$ & \tabularnewline
        \hline 
        $2$ & $\frac{2\pi}{3a}\left(0,0,0\right)$ & $0$ & $0$ & $0$ & $1$ & $\uparrow$ & $0$ & \multirow{2}{*}{$1.5191321$}\tabularnewline
        \cline{1-8} 
        $3$ & $\frac{2\pi}{3a}\left(0,0,0\right)$ & $0$ & $0$ & $0$ & $1$ & $\downarrow$ & $0$ & \tabularnewline
        \hline 
        $4$ & $\frac{2\pi}{3a}\left(1,0,0\right)$ & $1$ & $0$ & $0$ & $0$ & $\uparrow$ & $1$ & \multirow{2}{*}{$-0.5309026$}\tabularnewline
        \cline{1-8} 
        $5$ & $\frac{2\pi}{3a}\left(1,0,0\right)$ & $1$ & $0$ & $0$ & $0$ & $\downarrow$ & $1$ & \tabularnewline
        \hline 
        $6$ & $\frac{2\pi}{3a}\left(1,0,0\right)$ & $1$ & $0$ & $0$ & $1$ & $\uparrow$ & $0$ & \multirow{2}{*}{$0.6484053$}\tabularnewline
        \cline{1-8}  
        $7$ & $\frac{2\pi}{3a}\left(1,0,0\right)$ & $1$ & $0$ & $0$ & $1$ & $\downarrow$ & $0$ & \tabularnewline
        \hline 
        $8$ & $\frac{2\pi}{3a}\left(-1,0,0\right)$ & $-1$ & $0$ & $0$ & $0$ & $\uparrow$ & $1$ & \multirow{2}{*}{$-0.5309026$}\tabularnewline
        \cline{1-8} 
        $9$ & $\frac{2\pi}{3a}\left(-1,0,0\right)$ & $-1$ & $0$ & $0$ & $0$ & $\downarrow$ & $1$ & \tabularnewline
        \hline 
        $10$ & $\frac{2\pi}{3a}\left(-1,0,0\right)$ & $-1$ & $0$ & $0$ & $1$ & $\uparrow$ & $0$ & \multirow{2}{*}{$0.6484053$}\tabularnewline
        \cline{1-8} 
        $11$ & $\frac{2\pi}{3a}\left(-1,0,0\right)$ & $-1$ & $0$ & $0$ & $1$ & $\downarrow$ & $0$ & \tabularnewline
        \end{tabular}
    \end{ruledtabular}
\end{table}
\endgroup

\subsection{Unitary Coupled Cluster ansatz for periodic systems}

The molecular UCCSD ansatz can be easily relabelled to obtain the UCCSD-PBC ansatz for a localized supercell using the mapping $q_R(\mathbf{R},p,\sigma)$ to substitute the composite indices into the usual cluster operator expressions. 
For example, the cluster operators for closed shell configurations take the form
\begin{gather}
\hat{T}_{1}=\sum'_{\mathbf{R}_{p}\mathbf{R}_{q}}\sum'_{pq}t_{\mathbf{R}_{q}q}^{\mathbf{R}_{p}p}\sum_{\sigma}\hat{a}_{\mathbf{R}_{p}p\sigma}^{\dagger}\hat{a}_{\mathbf{R}_{q}q\sigma}
\\
\hat{T}_{2}=  \frac{1}{2} \sum'_{\mathbf{R}_{p}\mathbf{R}_{q}\mathbf{R}_{r}\mathbf{R}_{s}}\sum'_{pqrs}t_{\mathbf{R}_{q}q\mathbf{R}_{s}s}^{\mathbf{R}_{p}p\mathbf{R}_{r}r}\sum_{\sigma\sigma'}\hat{a}_{\mathbf{R}_{p}p\sigma}^{\dagger}\hat{a}_{\mathbf{R}_{q}q\sigma}\hat{a}_{\mathbf{R}_{r}r\sigma'}^{\dagger}\hat{a}_{\mathbf{R}_{s}s\sigma'}\end{gather}
where $t_{\mathbf{R}_{q}q}^{\mathbf{R}_{p}p}$ and $t_{\mathbf{R}_{q}q\mathbf{R}_{s}s}^{\mathbf{R}_{p}p\mathbf{R}_{r}r}$ are the cluster amplitudes corresponding to one- and two-electron excitations, respectively, and the prime symbol indicates that the summation is over the occupied to unoccupied excitations only. 
Since the system has translational invariance, for any $\mathbf{D}$ lattice translation, $t_{\mathbf{R}_{q}q}^{\mathbf{R}_{p}p} = t_{\mathbf{R}_{q}+\mathbf{D},q}^{\mathbf{R}_{p}+\mathbf{D},p}$ and $t_{\mathbf{R}_{q}q\mathbf{R}_{s}s}^{\mathbf{R}_{p}p\mathbf{R}_{r}r} = t_{\mathbf{R}_{q}+\mathbf{D}, q,\mathbf{R}_{s}+\mathbf{D}, s}^{\mathbf{R}_{p}+\mathbf{D}, p,\mathbf{R}_{r}+\mathbf{D}, r} = t_{\mathbf{R}_{s}s\mathbf{R}_{q}q}^{\mathbf{R}_{r}r\mathbf{R}_{p}p}$, which can be used to reduce the number of variational parameters in the VQE optimization. 
The cluster operators in the momentum space after the transformation in Equation \ref{eq:expansion} take the form
\begin{gather}
\hat{T}_{1}=\sum'_{\mathbf{k}_{p}\mathbf{k}_{q}}\sum'_{pq}t_{\mathbf{k}_{q}q}^{\mathbf{k}_{p}p}\sum_{\sigma}\hat{c}_{\mathbf{k}_{p}p\sigma}^{\dagger}\hat{c}_{\mathbf{k}_{q}q\sigma},
\\
\hat{T}_{2}= \frac{1}{2} \sum'_{\mathbf{k}_{p}\mathbf{k}_{r}\mathbf{k}_{s}\mathbf{k}_{q}}\sum'_{pqrs}t_{\mathbf{k}_{q}q\mathbf{k}_{s}s}^{\mathbf{k}_{p}p\mathbf{k}_{r}r}\sum_{\sigma\sigma'}\hat{c}_{\mathbf{k}_{p}p\sigma}^{\dagger}\hat{c}_{\mathbf{k}_{q}q\sigma}\hat{c}_{\mathbf{k}_{r}r\sigma'}^{\dagger}\hat{c}_{\mathbf{k}_{s}s\sigma'},
\end{gather}
where the prime symbol indicates the summation is over the set of occupied to unoccupied excitations and only for k-points which satisfy the momentum conservation law \cite{chan_ucc_pbc}. 
Consequently, the number of cluster amplitudes and excitations is reduced compared to the real space supercell case. 
The simplifications derived from crystal momentum conservation lead to a scaling of $N_{occ}^2N_{virt}^2L^3$ terms in the cluster operator, where $N_{occ}$ ($N_{virt}$) corresponds to the number of occupied (virtual) orbitals in the primitive cell.

In order to map the UCCSD-PBC ansatz to quantum circuits, the unitary coupled cluster operator is Trotterized and written as a product of Pauli exponentials by using the Jordan--Wigner transformation: 
\begin{equation}
e^{\hat{T}-\hat{T}^{\dagger}}\approx\prod_{m}\prod_{n}e^{i\theta_{m}P_{m,n}}
\label{trotterization}
\end{equation}
where $\theta_m$ are independent real parameters and $P_{m,n}$ are the corresponding Pauli strings. 
In the case of momentum space representation, $\hat{H}_K$, the Fourier transformation Equation \ref{eq:expansion} introduces complex orbitals, which in turn result in the presence of complex cluster amplitudes in the ansatz. 
One way to avoid the technical difficulties originating from complex amplitudes is the K2G approach discussed in Ref. \cite{liu_pbc}, which transforms the complex crystal orbitals to real $\Gamma$-HF orbitals of an equivalent supercell. 
In order to retain the favorable scaling properties of the momentum representation we directly map the excitations with complex amplitudes to quantum circuits. 
If the complex cluster amplitudes are written as
\begin{gather*}
t_{\mathbf{k}_{q}q}^{\mathbf{k}_{p}p}=u_{\mathbf{k}_{q}q}^{\mathbf{k}_{p}p}+iv_{\mathbf{k}_{q}q}^{\mathbf{k}_{p}p},
\\
t_{\mathbf{k}_{q}q\mathbf{k}_{s}s}^{\mathbf{k}_{p}p\mathbf{k}_{r}r}=u_{\mathbf{k}_{q}q\mathbf{k}_{s}s}^{\mathbf{k}_{p}p\mathbf{k}_{r}r}+iv_{\mathbf{k}_{q}q\mathbf{k}_{s}s}^{\mathbf{k}_{p}p\mathbf{k}_{r}r}
\end{gather*} then the anti-symmetrized cluster operators take the form of
\begin{widetext}
\begin{gather*}
\hat{T}_1-\hat{T}_1^{\dagger} = 
\sum'_{\substack{\mathbf{k}_{p}\mathbf{k}_{q}\\pq}} u_{\mathbf{k}_{q}q}^{\mathbf{k}_{p}p}
 \sum_{\sigma}\left(
\hat{c}_{\mathbf{k}_{p}p\sigma}^{\dagger}\hat{c}_{\mathbf{k}_{q}q\sigma}- \hat{c}_{\mathbf{k}_{q}q\sigma}^{\dagger}\hat{c}_{\mathbf{k}_{p}p\sigma}
\right) + 
i v_{\mathbf{k}_{q}q}^{\mathbf{k}_{p}p}
 \sum_{\sigma}\left(
\hat{c}_{\mathbf{k}_{p}p\sigma}^{\dagger}\hat{c}_{\mathbf{k}_{q}q\sigma}+ \hat{c}_{\mathbf{k}_{q}q\sigma}^{\dagger}\hat{c}_{\mathbf{k}_{p}p\sigma}
\right),
\\
\hat{T}_2-\hat{T}_2^{\dagger} = \frac{1}{2}
\sum'_{\substack{\mathbf{k}_{p}\mathbf{k}_{r}\mathbf{k}_{s}\mathbf{k}_{q}\\pqrs}}
u_{\mathbf{k}_{q}q\mathbf{k}_{s}s}^{\mathbf{k}_{p}p\mathbf{k}_{r}r}
\sum_{\sigma\sigma'}\left(
\hat{c}_{\mathbf{k}_{p}p\sigma}^{\dagger}\hat{c}_{\mathbf{k}_{q}q\sigma}\hat{c}_{\mathbf{k}_{r}r\sigma'}^{\dagger}\hat{c}_{\mathbf{k}_{s}s\sigma'}-
\hat{c}_{\mathbf{k}_{s}s\sigma'}^{\dagger}\hat{c}_{\mathbf{k}_{r}r\sigma'}\hat{c}_{\mathbf{k}_{q}q\sigma}^{\dagger}\hat{c}_{\mathbf{k}_{p}p\sigma}
\right) + \\ +
i v_{\mathbf{k}_{q}q\mathbf{k}_{s}s}^{\mathbf{k}_{p}p\mathbf{k}_{r}r}
\sum_{\sigma\sigma'}\left(
\hat{c}_{\mathbf{k}_{p}p\sigma}^{\dagger}\hat{c}_{\mathbf{k}_{q}q\sigma}\hat{c}_{\mathbf{k}_{r}r\sigma'}^{\dagger}\hat{c}_{\mathbf{k}_{s}s\sigma'}+
\hat{c}_{\mathbf{k}_{s}s\sigma'}^{\dagger}\hat{c}_{\mathbf{k}_{r}r\sigma'}\hat{c}_{\mathbf{k}_{q}q\sigma}^{\dagger}\hat{c}_{\mathbf{k}_{p}p\sigma}
\right)
\end{gather*}
\end{widetext} and the imaginary excitation operators are mapped to Pauli strings according to the Jordan--Wigner encoding,
\begin{widetext}
\begin{equation}
\label{eq:jw_one_re}
\hat{c}_{A}^{\dagger}\hat{c}_{I}-\hat{c}_{I}^{\dagger}\hat{c}_{A} = \frac{i}{2} \bigotimes_{M=q+1}^{A-1} Z_M \left( Y_I X_A - X_I Y_A \right),
\end{equation}
\begin{equation}
\label{eq:jw_one_im}
i \left( \hat{c}_{A}^{\dagger}\hat{c}_{I}+\hat{c}_{I}^{\dagger}\hat{c}_{A} \right) = \frac{i}{2} \bigotimes_{M=q+1}^{A-1} Z_M \left( X_I X_A + Y_I Y_A \right)
\end{equation},
\begin{multline}
\label{eq:jw_two_re}
\hat{c}_{A}^{\dagger} \hat{c}_{I} \hat{c}_{B}^{\dagger} \hat{c}_{J} - \hat{c}_{J}^{\dagger} \hat{c}_{B} \hat{c}_{I}^{\dagger} \hat{c}_{A} = \frac{i}{8} \bigotimes_{P=B+1}^{A-1} Z_P \bigotimes_{Q=J+1}^{I-1} Z_Q \left( Y_J Y_I Y_B X_A + X_J Y_I X_B X_A + Y_J Y_I X_B Y_A + Y_J X_I X_B X_A  - \right. \\ \left. Y_J X_I Y_B Y_A - X_J X_I X_B Y_A - X_J Y_I Y_B Y_A - X_J X_I Y_B X_A \right),
\end{multline} 
\begin{multline}
\label{eq:jw_two_im}
i\left(\hat{c}_{A}^{\dagger}\hat{c}_{I}\hat{c}_{B}^{\dagger}\hat{c}_{J}+\hat{c}_{J}^{\dagger}\hat{c}_{B}\hat{c}_{I}^{\dagger}\hat{c}_{A}\right) = \frac{i}{8} \bigotimes_{P=B+1}^{A-1} Z_P \bigotimes_{Q=J+1}^{I-1} Z_Q  \left( Y_J Y_I Y_B Y_A + X_J X_I X_B X_A + X_J Y_I X_B Y_A  + Y_J X_I Y_B X_A + \right. \\ \left.  X_J Y_I Y_B X_A + Y_J X_I X_B Y_A  - Y_J Y_I X_B X_A - X_J X_I Y_B Y_A \right), 
\end{multline}
\end{widetext}
where $A > B$  and $I > J$ are the mapped indices via $q_M(\mathbf{k},p,\sigma)$. Consequently, in the momentum representation, $\theta_m$ in Equation \ref{trotterization} corresponds to the real and imaginary parts of the cluster amplitudes separately with the Pauli strings in Equations 
\ref{eq:jw_one_re}, \ref{eq:jw_two_re}, and \ref{eq:jw_one_im}, \ref{eq:jw_two_im}, respectively. We note that the Pauli sub-terms of the real part of the cluster operators have an odd number of Pauli-$Y$s, whereas the sub-terms of the imaginary part have an even number of Pauli-$Y$s. The imaginary normalisation factor $\frac{i}{8}$ in equations \ref{eq:jw_one_re} and \ref{eq:jw_two_re} makes the entire expression real in contrast to equations \ref{eq:jw_one_im} and \ref{eq:jw_two_im} which are imaginary.
To further reduce the number of independent parameters one can use $t_{\mathbf{k}_{q}q}^{\mathbf{k}_{p}p}=t_{-\mathbf{k}_{q}q}^{-\mathbf{k}_{p}p*}$ and $t_{\mathbf{k}_{q}q\mathbf{k}_{s}s}^{\mathbf{k}_{p}p\mathbf{k}_{r}r}=\left(t_{-\mathbf{k}_{q}q-\mathbf{k}_{s}s}^{-\mathbf{k}_{p}p-\mathbf{k}_{r}r}\right)^{*}=t_{\mathbf{k}_{s}s\mathbf{k}_{q}q}^{\mathbf{k}_{r}r\mathbf{k}_{p}p}$. 
In the Appendix Tables \ref{tab:excitations_H2k2}, \ref{tab:excitations_H2k3} show an explicit list of the cluster operators in the order in which they are multiplied in the product in Equation \ref{trotterization}. In our calculations if it is not stated we did not enforce the symmetry constraints for the cluster amplitudes. 
Although enforcing the symmetry constraints is exact in CCSD, the Trotterization in Equation \ref{trotterization} does not in general preserve the equivalence of the parameters.

\subsection{Subspace Expansion with Translation operator}

The direct application of the UCCSD-PBC ansatz for strongly coupled systems may require a large number of k-points (Figure \ref{fig:resources}). 
In order to reduce the quickly growing circuit depth and the number of qubits necessary for a VQE calculation, we compute the ground state energy of the $\hat{H}_R$ via a subspace expansion with lattice translational operators. 
We refer to this method as TransQSE which provides an incremental way to tune between accuracy and feasibility of the VQE. 
The method can be viewed from the perspective of QSE, i.e. the subspace is constructed via lattice translational operators, or alternatively, can be viewed as a special case of non-orthogonal VQE \cite{Huggins_2020}. 
In what follows we discuss the TransQSE for linear chains.

Instead of the full UCCSD-PBC ansatz, a linear combination of translated and limited UCCSD ans\"{a}tze is introduced. 
The limited UCCSD ans\"{a}tze include excitations only in a $W<L$ number of primitive cells. In a 1D system such an ansatz can be constructed as
\begin{equation}
\ket{\Psi(\boldsymbol{\theta})}=\sum_{m = 0}^{L/\Delta - 1}c_{m}\hat{\Lambda}_{\Delta}^{m}\ket{\Psi_{W}(\boldsymbol{\theta})}
\end{equation}
where $c_{m}$ are complex coefficients, $\ket{\Psi_{W}(\boldsymbol{\theta})}$ is a UCCSD ansatz on a $W$ number of consecutive primitive cells and $\hat{\Lambda}_{\Delta}$ is a cyclic translation operator for the ansatz, which translates any state $\ket{\mathbf{R}}$ anchored at $\mathbf{R}$ to $\ket{\mathbf{R}+\mathbf{a}_1\Delta}=\hat{\Lambda}_{\Delta}\ket{\mathbf{R}}$.  
All other translated ans\"{a}tze can be obtained by subsequently applying the translation operator and for the supercell size $L=L_1$, $\hat{\Lambda}_{\Delta}^{L/\Delta} = \hat{I}$. 
The resulting subspace matrices are circulants, therefore the eigenvalues can be explicitly expressed. If for example $L/\Delta$ is odd
\begin{widetext}
\begin{equation}
E_{k_j}(\boldsymbol{\theta}) = \frac{\bra{\Psi_W(\boldsymbol{\theta})}\hat{H}_R\ket{\Psi_W(\boldsymbol{\theta})} + 2 \sum_{m=1}^{ \frac{L/\Delta-1}{2}}  \Re(\bra{\Psi_W(\boldsymbol{\theta})}\hat{H}_R\hat{\Lambda}^m_{\Delta}\ket{\Psi_W(\boldsymbol{\theta})} e^{i k_j m})}{1 + 2 \sum_{m=1}^{ \frac{L/\Delta-1}{2}} \Re( \bra{\Psi_W(\boldsymbol{\theta})}\hat{\Lambda}^m_{\Delta}\ket{\Psi_W(\boldsymbol{\theta})}e^{i k_j m})} 
\end{equation}
\end{widetext}
where $k_j = \frac{2\pi j}{L/\Delta}$ with $0\leq j<L/\Delta $ integers. The ground state energy in the VQE procedure is determined as
\begin{equation*}
E_0 = \min_{\boldsymbol{\theta}} \min_{k_j} E_{k_j}(\boldsymbol{\theta}). 
\end{equation*}
While the $\hat{H}_R$ operates on the Fock space of the entire supercell, the required qubits are determined by the qubit-region where the ansatz introduces entanglement. 
The size of the entanglement range has an upper bound at $2\times W \times2N$ qubits. 
The indices that belong to the non-entangled part of the ansatz can be contracted out from the $\hat{H}_R$ with the reference state; that yields  
\begin{multline*}
\bra{\Psi_{W}(\boldsymbol{\theta})}\hat{H}_R\hat{\Lambda}_{\Delta}^{m}\ket{\Psi_{W}(\boldsymbol{\theta})}= \\ \sum_{ij}h_{ij}^{W,\Delta,m}\Gamma_{ij}^{W,\Delta,m}(\boldsymbol{\theta})+\sum_{ijkl}h_{ijkl}^{W,\Delta,m}\Gamma_{ijkl}^{W,\Delta,m}(\boldsymbol{\theta})
\end{multline*} 
where the summation is over the entangled range, maximum $2\times W \times2N$ large, and $h_{ij}^{W,\Delta,m}$,$\Gamma_{ij}^{W,\Delta,m}(\boldsymbol{\theta})$,
$h_{ijkl}^{W,\Delta,m}$, $\Gamma_{ijkl}^{W,\Delta,m}(\boldsymbol{\theta})$ are the contracted Hamiltonian coefficients and the corresponding operator expectation values.  
Furthermore, if $m, m' > W$ then $\Gamma_{ij}^{W,\Delta,m}(\boldsymbol{\theta})=\Gamma_{ij}^{W,\Delta,m'}(\boldsymbol{\theta})$ and  $\Gamma_{ijkl}^{W,\Delta,m}(\boldsymbol{\theta})=\Gamma_{ijkl}^{W,\Delta,m'}(\boldsymbol{\theta})$
and also $\braket{\Psi_W(\boldsymbol{\theta})}{\hat{\Lambda}^m_{\Delta}\Psi_W(\boldsymbol{\theta})} = \braket{\Psi(\boldsymbol{\theta})}{\hat{\Lambda}^{m'}_{\Delta}\Psi(\boldsymbol{\theta})}$, which reduce the overall number of measurements. 
With the strategies discussed in \cite{Huggins_2020} the matrix elements can be mapped to circuits with maximum $4\times W \times2N+1$ qubits. 

\section{Results and Discussion}

\begin{figure}[ht]
\begin{center}
\includegraphics[width=\linewidth]{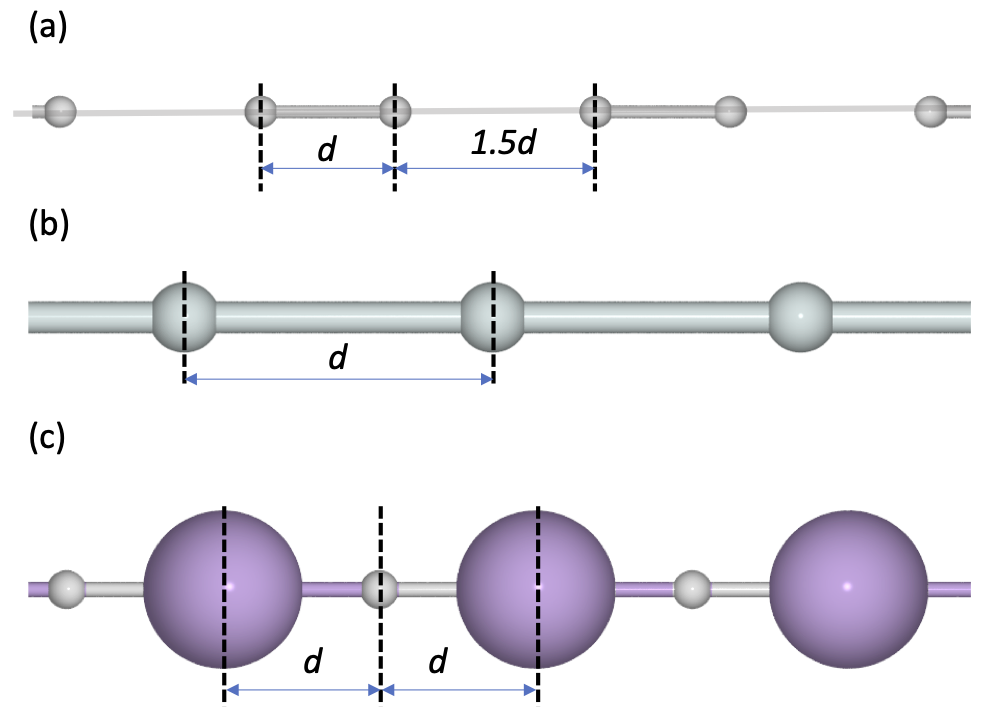}
\caption{Chain geometries with (a) hydrogen molecules, (b) helium atoms, (c) 1D lithium hydride molecules. 
}
\label{fig:geometries}
\end{center}
\end{figure}

\begin{figure}[ht]
\begin{center}
\includegraphics[width=\linewidth]{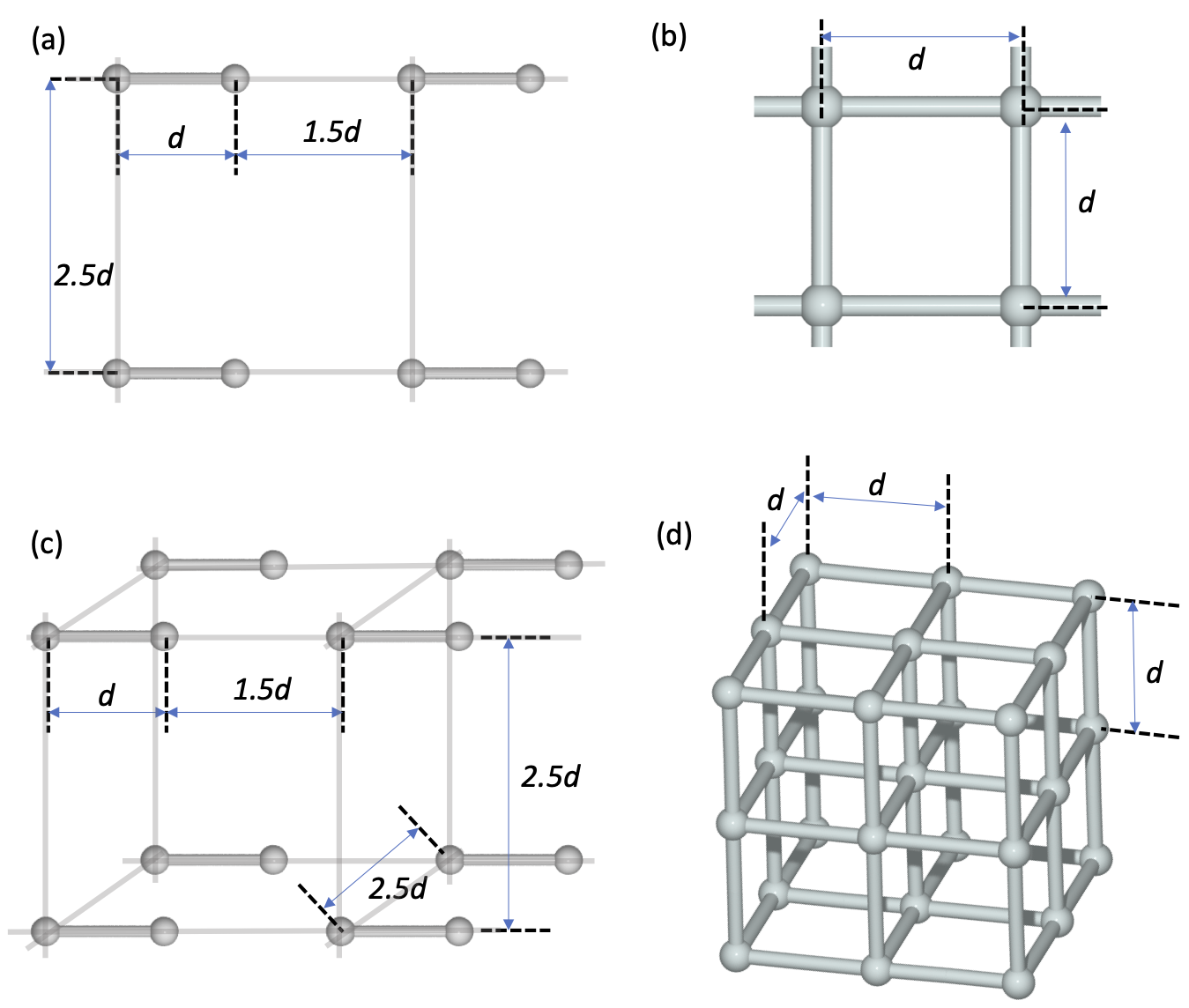}
\caption{Geometries for higher dimensional calculations with (a) hydrogen molecular lattice,
(b) helium atomic lattice,
(c) hydrogen molecular cube,
and (d) helium atomic cube.}
\label{fig:geometries_2d3d}
\end{center}
\end{figure}

In what follows we first demonstrate the discussed strategies on the hydrogen chain in detail and then show results for helium and lithium hydride chains and the 2D and 3D cubic forms of hydrogen and helium. 
For the chain models, we have performed simulations up to 5-k points ($L=2,3,4,5$) for a range of geometrical parameters. 
The exact geometries are described in Figure \ref{fig:geometries} for the 1D models and Figure \ref{fig:geometries_2d3d} for the 2D and 3D models. 
In order to obtain the Hamiltonian ($\hat{H}_K$) in the second quantized form, we performed 1D Restricted Hartree--Fock (RHF) calculations in the momentum space with PySCF that provided the nuclear repulsion energy, the Madelung correction and the one- and two electron integrals in the crystal orbitals. 
To compute the integrals, the Mixed Density Fitting method was used. To compute the localized form of the Hamiltonian ($\hat{H}_R$) we constructed the Wannier orbitals from the crystal orbitals with Equation \ref{eq:expansion}, and transformed the integrals accordingly. 
For reference, we also constructed the supercell geometries and performed $\Gamma$ point 1D RHF simulations to calculate FCI and CCSD energies and also built the second quantized Hamiltonian ($\hat{H}_\Gamma$) with the canonical orbitals and performed traditional UCCSD calculations. 
All of our results in this paper involve closed shell systems to reduce the computational cost of the calculations, but they can be extended to open shell systems. In the case of the hydrogen chains, we favoured the closed shell solution by applying a Peierls-type distortion that leads to two different H--H bond lengths in alternating positions along the chain, effectively transforming the system into a chain of H$_{2}$ molecules. We use our quantum chemistry package EUMEN for the quantum simulations, combined with some functions of the OpenFermion framework \cite{mcclean2017openfermion}, the state-vector simulator Qulacs \cite{Qulacs}, together with the retargetable $\ensuremath{\mathsf{t}|\mathsf{ket}\rangle}$ compiler\cite{Sivarajah_2020} for circuit optimization. The VQE calculations were performed with Scipy's L-BFGS-B minimizer\cite{2020SciPy-NMeth}.
In the calculations, for the hydrogen lattices we used STO-3G and for the helium lattices we used 6-31G basis sets. 
The LiH calculations were performed with a GTH-SZV basis set and GTH-HF pseudopotentials that replace the Li 1$s^2$ shell. 
In order to consistently compare the CCSD and UCCSD-PBC cluster amplitudes, for each amplitude provided by PySCF we added the corresponding excitations to the cluster operator. We constructed the cluster operators similarly for $\hat{H}_R$ and $\hat{H}_\Gamma$. 
In the Appendix Table \ref{tab:excitations_H2k3} for $L=3$ case we explicitly list the ordered excitations that were used in this paper. 

\subsection{Periodic {H$_2$} and the case {$L=2$}}

The number of terms in the different representations of the Hamiltonian significantly differ. 
For example, for the $L=2$, $\mathrm{H}_2$, $d=\SI{0.75}{\angstrom}$ case, for $\hat{H}_K$, $\hat{H}_R$ and $\hat{H}_\Gamma$, there are $97$, $185$ and $157$ unique normal ordered terms, respectively, with coefficient larger than $10^{-8}$ Ha magnitude. 
However, the number of reduced density matrix elements to be determined via measurements is lower if one considers the equivalency of some of the matrix elements within the closed shell UCCSD-PBC ansatz. 
In particular, $\hat{H}_R$ possesses easily exploitable translational symmetry, which significantly reduces the number of measurements. 
Similarly, the cluster operator is impacted by the choice of the Hamiltonian representation and the number of variational parameters in the UCCSD ansatz can be reduced by considering existing symmetries in the system. 
Since the expansion coefficients in Equation \ref{eq:expansion} are real for $L=2$, the introduction of complex amplitudes for the $\hat{H}_K$ can be avoided. 
The number of excitation operators in $\hat{T}$ are $24$, $44$ and $44$ for $\hat{H}_K$, $\hat{H}_R$ and $\hat{H}_\Gamma$, respectively and the number of cluster amplitudes are $10$, $20$ and $20$, respectively. 

In addition to the translational symmetry, the internal symmetries of the primitive cell can also be exploited to reduce the quantum resources needed to perform a quantum simulation. 
Although our closed shell UCCSD-PBC ansatz in the momentum space has $10$ variational parameters without the imaginary part of the amplitudes, but as it is shown in Table \ref{tab:cluster_amplitudes_H2k2}, one can omit the single excitations and the number of variational parameters eventually can be reduced to $6$ due to the redundancy of excitation operators (See also Table \ref{tab:excitations_H2k2_4qb} in the Appendix). We note that the Trotterized ansatz depends on the order of excitations, and the equivalency of some amplitudes breaks; for example we found that $t^{0,1,1,1}_{0,0,1,0} \neq t^{1,1,0,1}_{1,0,0,0}$ for the UCCSD-PBC. In general optimizing the ordering of the excitations in the cluster operators can lower the circuit depth \cite{hastings2014improving}. 

Furthermore, symmetries, in particular the translational symmetry, can be used to reduce the required number of qubits as well. It has been shown previously that some point group symmetries of molecular systems correspond to $Z_2$ symmetries \cite{setia2019reducing}. 
This idea is extended further to taper qubits for periodic systems by noting that space groups are used to characterize the symmetries of infinite periodic systems, correspond to point groups in reciprocal space. For example, without considering symmetries, the $L=2$ $\mathrm{H}_2$ system is simulated on $8$ qubits, but we found that the quantum circuits can be reduced to 4 qubits by tapering off qubits corresponding to $Z_2$ symmetries \cite{bravyi2017tapering}.
By using the indexing in Table \ref{tab:encodings_H2k2}, we identified the following symmetries in the $\hat{H}_K$ of $L=2$ $\mathrm{H}_2$ system: the number symmetry of spin-up electrons $\left( Z_0 Z_2 Z_4 Z_6 \right)$, number symmetry of spin-down electrons $\left( Z_1 Z_3 Z_5 Z_7 \right)$, the internal symmetry of the $\mathrm{H}_2$ molecule $\left( Z_2 Z_3 Z_6 Z_7 \right)$ and the translational symmetry across the supercell $\left( Z_4 Z_5 Z_6 Z_7 \right)$. We use these $Z_2$ symmetries in combination with functionality from the OpenFermion framework to taper off qubits. The derivation is elaborated in the Appendix and the optimized amplitudes obtained with the 4-qubit system are listed in Table \ref{tab:excitations_H2k2_4qb}.

\begingroup
\squeezetable
\begin{table}
    \caption{Cluster amplitudes for the hydrogen chain $d=\SI{0.75}{\angstrom}$, $L=2$ case in momentum space.}
    \label{tab:cluster_amplitudes_H2k2}
    \begin{ruledtabular}
        \begin{tabular}{cddd}
        \multicolumn{1}{c}{$t_{\mathbf{k}_{q},q}^{\mathbf{k}_{p},p}$ , $t_{\mathbf{k}_{q},q,\mathbf{k}_{s},s}^{\mathbf{k}_{p},p,\mathbf{k}_{r},r}$} \footnote{Since it is a 1D system, the $\mathbf{k}$ is represented by the number $k^{(1)}$ in the amplitude symbol.}
        & \multicolumn{1}{c}{Initial}
        & \multicolumn{1}{c}{CCSD}
        & \multicolumn{1}{c}{UCCSD-PBC}
        \tabularnewline
        
        \hline
        $t_{0, 0}^{0, 1}$
        &        0
        &        0
        &        0
        \tabularnewline
        
        \hline
        $t_{1, 0}^{1, 1}$
        &        0
        &        0
        &        0
        \tabularnewline
        
        \hline
        $t_{0, 0, 0, 0}^{0, 1, 0, 1}$
        & -0.016154
        & -0.022288
        & -0.022966
        \tabularnewline
        
        \hline
        $t_{0, 0, 0, 0}^{1, 1, 1, 1}$
        & -0.056961
        & -0.082398
        & -0.081645
        \tabularnewline
        
        \hline
        $t_{0, 0, 1, 0}^{0, 1, 1, 1}$
        &  0.02126
        & 0.038047
        & 0.037889
        \tabularnewline
        
        \hline
        $t_{0, 0, 1, 0}^{1, 1, 0, 1}$       
        &   0.0395
        & 0.057676
        & 0.057322
        \tabularnewline
        
        \hline
        $t_{1, 0, 0, 0}^{0, 1, 1, 1}$\footnote{Redundant excitations.}
        &   0.0395
        & 0.057676
        & 0.057322
        \tabularnewline

        \hline
        $t_{1, 0, 0, 0}^{1, 1, 0, 1}$       $^{\text{b}}$
        &  0.02126
        & 0.038047
        & 0.037798
        \tabularnewline

        \hline
        $t_{1, 0, 1, 0}^{0, 1, 0, 1}$       
        & -0.034029
        & -0.041929
        & -0.04174
        \tabularnewline
        
        \hline
        $t_{1, 0, 1, 0}^{1, 1, 1, 1}$
        & -0.036581
        & -0.099775
        & -0.099503
        \tabularnewline
        
        \end{tabular}
    \end{ruledtabular}
\end{table}
\endgroup

\subsection{Chains with higher value of $L$}

\begin{figure}
\begin{center}
\includegraphics[width=\linewidth]{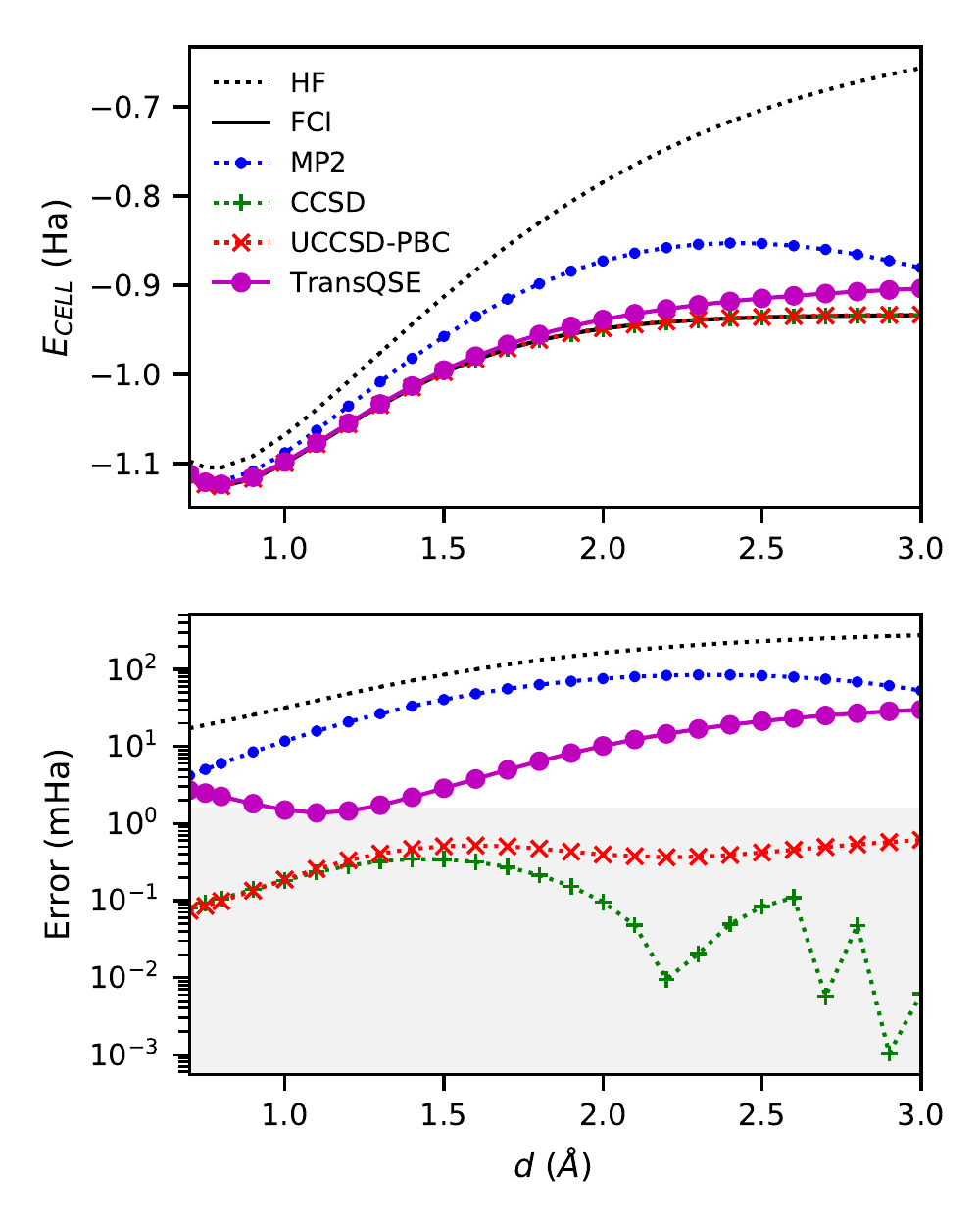}
\caption{Energy calculations for $L=3$ for hydrogen  chains. The shaded region is below the accuracy of $1.0$ kcal/mol = $1.6 \times 10^{-3}$ Ha. The HF, FCI, MP2 and CCSD energies were calculated from $\Gamma$-point computations of the supercell.
}
\label{fig:energies}
\end{center}
\end{figure}

In the momentum space representation, the $L = 3$ case demonstrates the occurrence of complex excitations which are converted to qubit operators according to Equations \ref{eq:jw_one_re} -- \ref{eq:jw_two_im}. In the VQE procedure the real and imaginary parts of the complex amplitudes are treated as independent parameters. The optimized complex amplitudes for H$_2$ are in Table \ref{tab:cluster_amplitudes_H2k3}, which shows that the UCCSD-PBC amplitudes are similar to the CCSD amplitudes. The TransQSE method for $L=2$ case is too simple, therefore we demonstrate this method in the $L=3$ case, with a restricted window of $W=2$ and $\Delta=1$. The relevant subspace matrix elements are
\begin{gather} 
h_{0}=\bra{\Psi_{W}(\boldsymbol{\theta})}H\ket{\Psi_{W}(\boldsymbol{\theta})},
\\
h_{1}=\bra{\Psi_{W}(\boldsymbol{\theta})}H\hat{\Lambda}_{\Delta}\ket{\Psi_{W}(\boldsymbol{\theta})},
\\
s_{1}=\bra{\Psi_{W}(\boldsymbol{\theta})}\hat{\Lambda}_{\Delta}\ket{\Psi_{W}(\boldsymbol{\theta})}.
\end{gather} 
With the strategies discussed in \cite{Huggins_2020} the subspace matrix elements can be mapped to circuits. In our example calculations we used a state-vector simulation to evaluate $h_{0}, h_{1}$ and $s_{1}$. Once the matrix elements are obtained for a given set of $\boldsymbol{\theta}$, the energy function 
\begin{equation}
E(\boldsymbol{\theta})=\min_{n=-1,0,1}\frac{h_{0}+2\Re (h_{1}e^{\frac{2\pi i}{3}n})}{1+2\Re (s_{1}e^{\frac{2\pi i}{3}n})}
\end{equation} 
is evaluated to find the minimum energy according to the VQE minimization. Figure \ref{fig:energies} shows the ground state energies for a range of geometrical parameters, as defined in Figure \ref{fig:geometries}. The initial value of the variational parameters were randomly generated with normal distribution with $\sigma_{N}=0.01$ and we found that the L-BFGS-B minimizer may not always find the lowest energy.  

\begin{figure}
\begin{center}
\includegraphics[width=\linewidth]{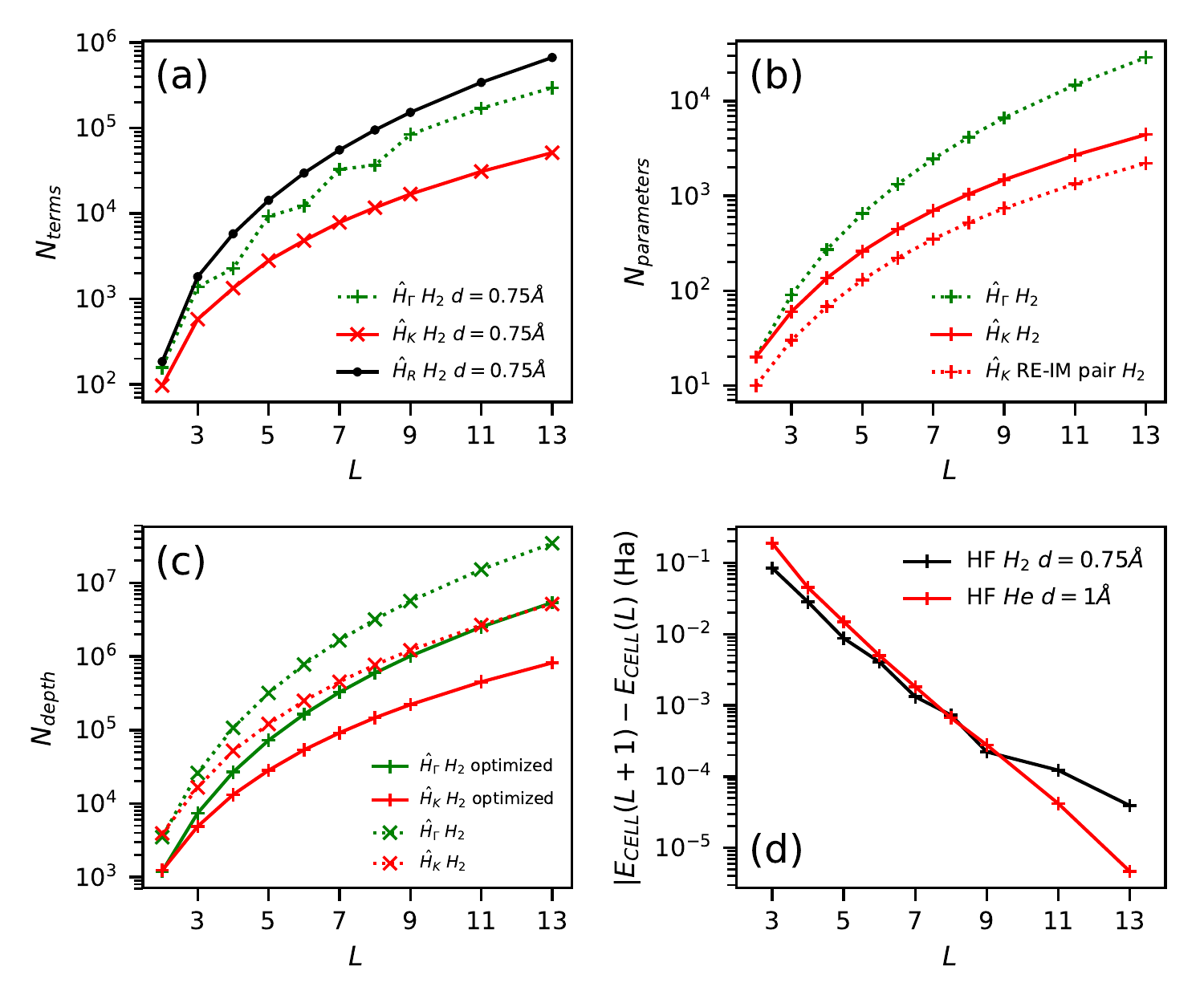}
\caption{ Computational resources needed for hydrogen and helium chains UCCSD-PBC calculation. The circuit optimization was performed with $\texttt{GuidedPauliSimp}$.}
\label{fig:resources}
\end{center}
\end{figure}

General comparisons for larger values of $L$ and for H$_2$, He and LiH chains (shown in Figure \ref{fig:geometries}) are summarised in Table \ref{tab:energy_comparisons}. 
We found that the single Trotter step UCCSD-PBC VQE procedure in different representations results in very similar energies, and also for the investigated systems the CCSD and UCCSD-PBC energies differ less than the chemical accuracy. 
Figure \ref{fig:energies} shows the comparison of CCSD and UCCSD-PBC energies for a range of geometrical parameters defined in Figure \ref{fig:geometries}. Figure \ref{fig:energies} and Table \ref{tab:energy_comparisons} also show the comparisons with the TransQSE energies with $W=2$ and $\Delta=1$. In the compressed chains the TransQSE energies are closer to FCI energies than in the stretched geometries. We also found that to reach the converged mean-field energies, (See Figure \ref{fig:resources}) values of $L$ in the order of $10$ or more are required. This is not unusual for periodic structures with strong couplings between their primitive cells. Figure \ref{fig:resources} shows the typical scaling behaviour of the resources to perform a full UCCSD-PBC method for increasing $L$. For example, to perform an $L=9$ simulation, the required circuit depths reach values of the order of $10^5$ and the number of terms reaches the order of $10^4$. We found that the $\hat{H}_K$ representation is the most favourable as the circuit depth and the number of terms are smaller than that of the other representations, however without significant reductions they are not suitable for NISQ architecture. The TransQSE provides a systematic approach to reduce the computational cost, as the quantum computational resources do not scale with $L$ but with $W$. For example, the TransQSE energy for $L=9$ with $W=2$ requires only $20$ parameters and a circuit depth similar to an $L=2$ UCCSD calculation.

\begin{figure}
\begin{center}
\includegraphics[width=\linewidth]{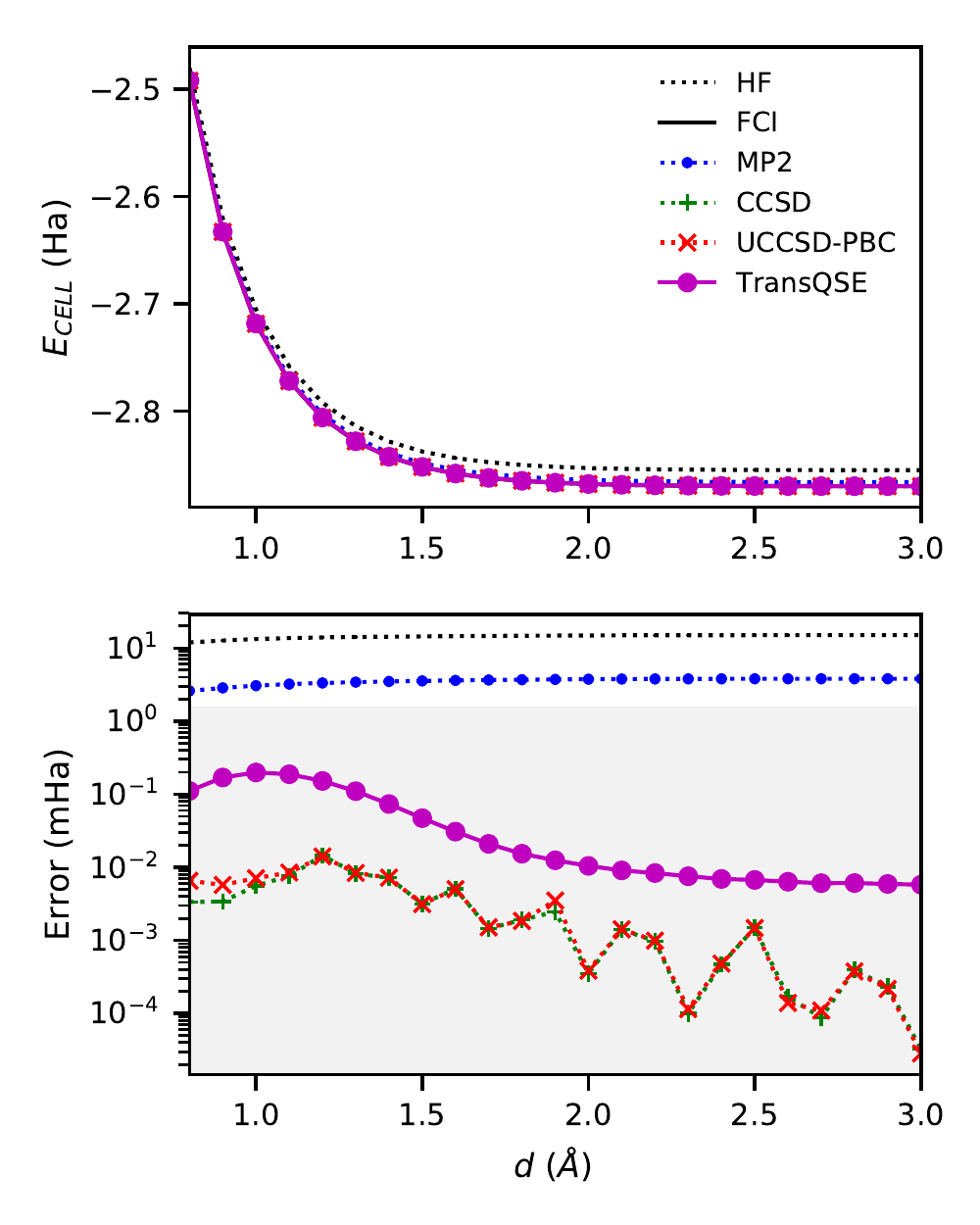}
\caption{Energy calculations for $L=3$ for helium  chains. The shaded region is below the accuracy of $1.0$ kcal/mol = $1.6 \times 10^{-3}$ Ha. The HF, FCI, MP2 and CCSD energies were calculated from $\Gamma$-point computations of the supercell.
}
\label{fig:energies_he_l3}
\end{center}
\end{figure}
\subsection{2- and 3-dimensional cubic hydrogen and helium}

We also extended the UCCSD-PBC approach to 2D and 3D systems. The computational scaling with the number of $L$ is similar to the 1D case, however $L_1$, $L_2$ and $L_3$ separately need to be increased to reach accurate mean-field energies. Therefore the overall computational costs increase even more rapidly compared to the 1D case.

To demonstrate that the higher dimensional VQE calculations return similar energies to those of classical counterparts as in the 1D cases, we consider simple square and cubic lattices for 2D and 3D cases, respectively. 
Here H$_{2}$ and He systems are focused on, and all the primitive cell settings including nuclear positions and atomic basis sets remain the same as the corresponding 1-dimensional systems.                                        
We sample a [2, 2] Monkhorst--Pack k-point mesh \cite{monkhorst_1976} in the xy plane ($L = 4$) for both 2- and 3-dimensional cases.              
The sampling along the z axis should be included for the regular 3-dimensional analysis, but the direct UCCSD calculations with $L=8$ are prohibitively expensive for simulators and devices available today.   

The resulting energies from the higher dimensional UCCSD-PBC calculations are shown in Table \ref{tab:energy_comparisons}.                    See the rows with $D = 2, 3$ and $L = 4$.                                       
The accuracy of all the higher dimensional UCCSD-PBC calculations is commonly better than $10^{-5}$ Ha, which is even better than chemical accuracy ($1.0$ kcal/mol = $1.6 \times 10^{-3}$ Ha) compared to the FCI results. 
Larger calculations including the 3D symmetric k-point sampling ($D = 3, L = 8$) will be considered in future work.

\begingroup
\begin{table*}
     \caption{Comparison of energy values (in Ha) calculated with different methods on the periodic systems considered in this work. D stands for the number of dimensions in the periodic model.}
    \label{tab:energy_comparisons}
    \begin{ruledtabular}
    \begin{tabular}{lccc|ccc|ccc}
        & 
        $d$ ($\SI{}{\angstrom}$) &
        $D$ &
        $L$ & 
        $E_{\textrm{FCI}}^{\Gamma}/L$ & 
        $E_{\textrm{CCSD}}^{\Gamma}/L$ & 
        $E_{\textrm{UCCSD-PBC}}^{\Gamma}/L$ & 
        $E_{\textrm{UCCSD-PBC}}^{R}$ & 
        $E_{\textrm{UCCSD-PBC}}^{K}$ & 
        $E_{\textrm{TransQSE-W(2)}}$ 
        \\
        \hline
        H$_{2}$ & 
        $0.75$ &
        1 & 
        2 &
        $-1.0414576078$ & 
        $-1.0414491762$ & 
        $-1.0414573244$ & 
        $-1.0414573310$ &
        $-1.0414573245$ &
        --
        \\
        H$_{2}$ & 
        $0.75$ &
        1 & 
        3 & 
        $-1.1232654641$ & 
        $-1.1231727367$ & 
        $-1.1231804970$ & 
        $-1.1231804857$ &
        $-1.1231804672$ &
        $-1.1207870419$
        \\ 
        H$_{2}$ & 
        $0.75$ &
        1 & 
        5 & 
        $-1.1051530344$ & 
        $-1.1050265451$ & 
        $-1.1050300023$ & 
        $-1.1050299871$ &
        $-1.1050299775$ &
        -- 
        \\ 
        H$_{2}$ & 
        $0.75$ &
        1 & 
        9 & 
        -- & 
        $-1.1019109184$ & 
        -- & 
        -- &
        -- &
        -- 
        \\ 
        \hline               
        H$_{2}$ & 
        $0.75$ &
        2 & 
        4 & 
        $-1.0080293562$ & 
        $-1.0080185429$ & 
        $-1.0080274642$ & 
        -- &
        $-1.0080272592$ &
        --
        \\ 
        \hline               
        H$_{2}$\footnote{A  [2, 2] k-point mesh in the xy plane is selected for this calculation.} & 
        $0.75$ &
        3 & 
        4 & 
        $-1.2012791803$ & 
        $-1.2012520442$ & 
        $-1.2012684209$ & 
        -- &
        $-1.2012686107$ &
        --
        \\
        \hline
        He & 
        $1.0$ &
        1 & 
        2 &
        $-2.5285938200$ & 
        $-2.5285812803$ & 
        $-2.5285798759$ & 
        $-2.5285798742$ &
        $-2.5285798759$ &
        --
        \\ 
        He & 
        $1.0$ &
        1 & 
        3 & 
        $-2.7187307339$ & 
        $-2.7187230508$ & 
        $-2.7187216204$ & 
        $-2.7187266239$ &
        $-2.7187266447$ &
        $-2.7185322012$  
        \\ 
        He & 
        $1.0$ &
        1 & 
        9 & 
        -- & 
        $-2.6848072550$ & 
        -- & 
        -- &
        -- &
        -- 
        \\ 
        \hline               
        He & 
        $1.0$ &
        2 & 
        4 & 
        $-2.2583587870$ & 
        $-2.2583498391$ & 
        $-2.2583476552$ & 
        -- &
        $-2.2583476554$ &
        --
        \\ 
        \hline               
        He$^{\text{a}}$ & 
        $1.0$ &
        3 & 
        4 & 
        $-2.9562190137$ & 
        $-2.9562164816$ & 
        $-2.9562162171$ & 
        -- &
        $-2.9562162173$ &
        --
        \\
        \hline               
        LiH & 
        $2.0$ &
        1 & 
        3 & 
        $-7.9828384941$ & 
        $-7.9824545358$ & 
        $-7.9825328540$ & 
        $-7.9825334480$ &
        $-7.9825334389$ &
        --
        \\
    \end{tabular}
    \end{ruledtabular}

\end{table*}
\endgroup

In conclusion, we have demonstrated VQE calculations for periodic systems based second-quantized Hamiltonians in localized and momentum space representations. 
Choosing different representations of the Hamiltonian impacts the circuit depth, the number of variational parameters and number of measurements significantly. 
We also found that the introduction of complex amplitudes requires additional attention and if the imaginary part of the cluster amplitudes is present the number of Pauli exponentials doubles. However the circuit depth of the UCCSD-PBC ansatz is found to be the most favorable in the momentum space representation.

The resource requirements for the full UCCSD-PBC ansatz become prohibitively expensive for calculations with large numbers of k-points, even on quantum computers. 
We found that by choosing a smaller supercell, consisting of only $W<L$ primitive cells for the entangled domain, TransQSE can trade accuracy effectively against the expensive scaling of the full UCCSD-PBC approach. The method can be useful for systems requiring a large number of k-points, where scalability of the method in $L$ is important. 

It is notable however that the discussed strategy only works for insulating systems, where the occupation of the orbitals do not change with different k-points. 
For metals, where band lines cross the Fermi level, the transformation described by Equation \ref{eq:expansion} mixes occupied and virtual orbitals and it is not straightforward to prepare the reference state. 
An alternative approach is to use approximation methods such as maximally localized Wannier orbitals that can generate a realistic localized model Hamiltonian, which opens up an avenue for more theoretical investigations \cite{Marzari_2012}. 
While the UCCSD-PBC ansatz includes single and double excitations, we may expect future methods selecting the most relevant excitations to reduce the computation cost further. 
The TransQSE method is also complementary to embedding methods \cite{pham2019periodic}, ADAPT-VQE \cite{adapt_vqe} and modified UCCGSD-based VQE methods \cite{lee_kupcc} for periodic systems \cite{liu_pbc}, and even can be extended for non-chemistry related ans\"{a}tze.

\begin{acknowledgments}
VW worked on initial investigations and theory during his 2-month internship program at Cambridge Quantum Computing. We appreciate useful discussions with Dr. Koji Hirano and Dr. Hideaki Sawada from Nippon Steel Corporation. The numerical simulations in this work were performed on Microsoft Azure Virtual Machines provided by the program Microsoft for Startups. 
\end{acknowledgments}


\bibliographystyle{apsrev}
\bibliography{references.bib} 

\clearpage
\onecolumngrid
\appendix{APPENDIX}
\section{Cluster Operators for Various Hydrogen Chains}
\begingroup
\squeezetable
\begin{center}
\begin{longtable}{lddd}
    \caption{Cluster amplitudes for the hydrogen chain     \label{tab:cluster_amplitudes_H2k3}
    $d=\SI{0.75}{\angstrom}$, $L=3$ case in momentum space.}
    
    \tabularnewline
    
    \multicolumn{1}{c}{$t_{\mathbf{k}_{q}q}^{\mathbf{k}_{p}p}$ / $t_{\mathbf{k}_{q}q\mathbf{k}_{s}s}^{\mathbf{k}_{p}p\mathbf{k}_{r}r}$}
    & \multicolumn{1}{c}{Initial}
    & \multicolumn{1}{c}{CCSD}
    & \multicolumn{1}{c}{UCCSD-VQE}
    
    \endfirsthead
    \endhead
    \endfoot
    \hline \hline
    \endlastfoot
    
    \hline \hline
    $\Re t_{0, 0}^{0, 1}$                   
    & 0.000000
    & 0.000000
    & 0.000000
    \tabularnewline
    
    \hline
    $\Im t_{0, 0}^{0, 1}$                   
    & 0.000000
    & 0.000000
    & 0.000000
    \tabularnewline
    
    \hline
    $\Re t_{1, 0}^{1, 1}$                   
    & 0.000000
    & -0.000003
    & -0.000009
    \tabularnewline
    
    \hline
    $\Im t_{1, 0}^{1, 1}$                   
    & 0.000000
    & 0.000194
    & 0.000628
    \tabularnewline
    
    \hline
    $\Re t_{-1, 0}^{-1, 1}$                 
    & 0.000000
    & -0.000003
    & -0.000009
    \tabularnewline
    
    \hline
    $\Im t_{-1, 0}^{-1, 1}$                 
    & 0.000000
    & -0.000194
    & -0.000626
    \tabularnewline
    
    \hline
    $\Re t_{0, 0, 0, 0}^{0, 1, 0, 1}$       
    & -0.010619
    & -0.015747
    & -0.015933
    \tabularnewline
    
    \hline
    $\Im t_{0, 0, 0, 0}^{0, 1, 0, 1}$       
    & 0.000000
    & 0.000000
    & 0.000000
    \tabularnewline
\end{longtable}
\end{center}
\endgroup

\begin{center}
    \begin{longtable}{>{\tiny}c>{\tiny}c>{\tiny}c>{\tiny}c>{\tiny}c>{\tiny}c}
    \caption{Cluster operators for the Hydrogen chain STO-3G, $L=2$ case in momentum space.}
    \label{tab:excitations_H2k2}
    \tabularnewline
    
    \multicolumn{1}{>{\tiny}c}{Amplitudes}
    & \multicolumn{1}{>{\tiny}c}{Excitations}
    & \multicolumn{1}{>{\tiny}c}{Parameters}
    & \multicolumn{1}{>{\tiny}c}{Reduced Parameters} 
    & \multicolumn{1}{>{\tiny}c}{Exponents}
    & \multicolumn{1}{>{\tiny}c}{Tapered Exponents}
    
    \endfirsthead
    \endhead
    \endfoot
    \hline \hline
    \endlastfoot
    
    \hline \hline
    
    $\Re t_{0, 0}^{0, 1}$                   
    & \makecell{$ c^{\dagger}_{0,1,\uparrow}c_{0,0,\uparrow}$\\$+ c^{\dagger}_{0,1,\downarrow}c_{0,0,\downarrow}$}
    & $\theta_{0}$                            
    & --                           
    & \makecell{$-0.5\mathrm{i} X_{ 0 }Z_{ 1 }Y_{ 2 }$$+0.5\mathrm{i} Y_{ 0 }Z_{ 1 }X_{ 2 }$\\$-0.5\mathrm{i} X_{ 1 }Z_{ 2 }Y_{ 3 }$$+0.5\mathrm{i} Y_{ 1 }Z_{ 2 }X_{ 3 }$}
    & --                           
    \tabularnewline
    
    \hline
    $\Im t_{0, 0}^{0, 1}$                   
    & \makecell{$\mathrm{i} c^{\dagger}_{0,1,\uparrow}c_{0,0,\uparrow}$\\$+\mathrm{i} c^{\dagger}_{0,1,\downarrow}c_{0,0,\downarrow}$}
    & $\theta_{1}$                            
    & --                           
    & \makecell{$0.5\mathrm{i} X_{ 0 }Z_{ 1 }X_{ 2 }$$+0.5\mathrm{i} Y_{ 0 }Z_{ 1 }Y_{ 2 }$\\$+0.5\mathrm{i} X_{ 1 }Z_{ 2 }X_{ 3 }$$+0.5\mathrm{i} Y_{ 1 }Z_{ 2 }Y_{ 3 }$}
    & --                           
    \tabularnewline
    
    \hline
    $\Re t_{1, 0}^{1, 1}$                   
    & \makecell{$ c^{\dagger}_{1,1,\uparrow}c_{1,0,\uparrow}$\\$+ c^{\dagger}_{1,1,\downarrow}c_{1,0,\downarrow}$}
    & $\theta_{2}$                            
    & --                           
    & \makecell{$-0.5\mathrm{i} X_{ 4 }Z_{ 5 }Y_{ 6 }$$+0.5\mathrm{i} Y_{ 4 }Z_{ 5 }X_{ 6 }$\\$-0.5\mathrm{i} X_{ 5 }Z_{ 6 }Y_{ 7 }$$+0.5\mathrm{i} Y_{ 5 }Z_{ 6 }X_{ 7 }$}
    & --                           
    \tabularnewline
    
    \hline
    $\Im t_{1, 0}^{1, 1}$                   
    & \makecell{$\mathrm{i} c^{\dagger}_{1,1,\uparrow}c_{1,0,\uparrow}$\\$+\mathrm{i} c^{\dagger}_{1,1,\downarrow}c_{1,0,\downarrow}$}
    & $\theta_{3}$                            
    & --                           
    & \makecell{$0.5\mathrm{i} X_{ 4 }Z_{ 5 }X_{ 6 }$$+0.5\mathrm{i} Y_{ 4 }Z_{ 5 }Y_{ 6 }$\\$+0.5\mathrm{i} X_{ 5 }Z_{ 6 }X_{ 7 }$$+0.5\mathrm{i} Y_{ 5 }Z_{ 6 }Y_{ 7 }$}
    & --                           
    \tabularnewline
\end{longtable}
\end{center}

\begingroup
\squeezetable

\begin{center}
    \begin{longtable}{>{\tiny}c>{\tiny}c>{\tiny}c>{\tiny}c}
    \caption{Cluster operators for the Hydrogen chain STO-3G, $L=3$ case in momentum space.}
    \label{tab:excitations_H2k3}

    \tabularnewline
    
    \multicolumn{1}{>{\tiny}c}{Amplitudes}
    & \multicolumn{1}{>{\tiny}c}{Excitations}
    & \multicolumn{1}{>{\tiny}c}{Parameters}
    & \multicolumn{1}{>{\tiny}c}{Exponents}
    
    \endfirsthead
    \endhead
    \endfoot
    \hline \hline
    \endlastfoot
    
    \hline \hline
    $\Re t_{0, 0}^{0, 1}$                   
    & \makecell{$ c^{\dagger}_{0,1,\uparrow}c_{0,0,\uparrow}$\\$+ c^{\dagger}_{0,1,\downarrow}c_{0,0,\downarrow}$}
    & $\theta_{0}$                            
    & \makecell{$-0.5\mathrm{i} X_{ 0 }Z_{ 1 }Y_{ 2 }$$+0.5\mathrm{i} Y_{ 0 }Z_{ 1 }X_{ 2 }$\\$-0.5\mathrm{i} X_{ 1 }Z_{ 2 }Y_{ 3 }$$+0.5\mathrm{i} Y_{ 1 }Z_{ 2 }X_{ 3 }$}
    \tabularnewline
    
    \hline
    $\Im t_{0, 0}^{0, 1}$                   
    & \makecell{$\mathrm{i} c^{\dagger}_{0,1,\uparrow}c_{0,0,\uparrow}$\\$+\mathrm{i} c^{\dagger}_{0,1,\downarrow}c_{0,0,\downarrow}$}
    & $\theta_{1}$                            
    & \makecell{$0.5\mathrm{i} X_{ 0 }Z_{ 1 }X_{ 2 }$$+0.5\mathrm{i} Y_{ 0 }Z_{ 1 }Y_{ 2 }$\\$+0.5\mathrm{i} X_{ 1 }Z_{ 2 }X_{ 3 }$$+0.5\mathrm{i} Y_{ 1 }Z_{ 2 }Y_{ 3 }$}
    \tabularnewline
    
    \hline
    $\Re t_{1, 0}^{1, 1}$                   
    & \makecell{$ c^{\dagger}_{1,1,\uparrow}c_{1,0,\uparrow}$\\$+ c^{\dagger}_{1,1,\downarrow}c_{1,0,\downarrow}$}
    & $\theta_{2}$                            
    & \makecell{$-0.5\mathrm{i} X_{ 4 }Z_{ 5 }Y_{ 6 }$$+0.5\mathrm{i} Y_{ 4 }Z_{ 5 }X_{ 6 }$\\$-0.5\mathrm{i} X_{ 5 }Z_{ 6 }Y_{ 7 }$$+0.5\mathrm{i} Y_{ 5 }Z_{ 6 }X_{ 7 }$}
    \tabularnewline
    
    \hline
    $\Im t_{1, 0}^{1, 1}$                   
    & \makecell{$\mathrm{i} c^{\dagger}_{1,1,\uparrow}c_{1,0,\uparrow}$\\$+\mathrm{i} c^{\dagger}_{1,1,\downarrow}c_{1,0,\downarrow}$}
    & $\theta_{3}$                            
    & \makecell{$0.5\mathrm{i} X_{ 4 }Z_{ 5 }X_{ 6 }$$+0.5\mathrm{i} Y_{ 4 }Z_{ 5 }Y_{ 6 }$\\$+0.5\mathrm{i} X_{ 5 }Z_{ 6 }X_{ 7 }$$+0.5\mathrm{i} Y_{ 5 }Z_{ 6 }Y_{ 7 }$}
    \tabularnewline
\end{longtable}
\end{center}
\endgroup

\clearpage
\section{Resource Reduction for Hydrogen Chain $L=2$, d=\SI{0.75}{\angstrom} using Point Group Symmetries}
By imposing Born-von Karman boundary conditions, a periodic linear system can be shown to be equivalent to a cyclic system. As a direct consequence, the translation operation on the L=2 hydrogen chain in real space is linked to a $C_2$ rotation in reciprocal space. Using this relation, we can analyze the periodicity of the crystal system using point group symmetries \cite{SolidOrbitalBook}. 

To find the $Z_2$ symmetries we need to first determine the symmetry operation in second quantization. We restate the method here for convenience \cite{Yen_2019, setia2019reducing}. For Abelian groups, the symmetry operation $\hat{S}$ acting on an N spin-orbital system can be defined as a unitary operation,
\begin{equation}
    \label{eq:unitary_transform}
    \hat{S} = \prod_{j=0}^{N-1} e^{-i \theta \hat{a}_j^{\dagger} \hat{a}_j},
\end{equation}
where $a_j$ is a Fermion operator acting on orbital $j$. The presence of the dagger (lack of dagger) represent the creation (annihilation) of an electron in spin-orbital $j$. The rotation parameter $\theta$ must necessarily be 0 or $\pi$. The number operator in Jordan-Wigner encoding is $\frac{1}{2} \left( \hat{I}_j - \hat{Z}_j \right)$, where $\hat{I}_j$ and $\hat{Z}_j$ are Pauli operators, and $j$ is the qubit index. The symmetry operator in Jordan-Wigner $\hat{S}^{JW}_j$ acting on arbitrary qubit $j$ is, 
\begin{equation}
    \label{eq:unitary_transform_j_jk}
    \hat{S}^{JW}_j = e^{-i \frac{\theta}{2} I_j} e^{i \frac{\theta}{2} Z_j}.
\end{equation}
For the case when $\theta$ is zero, $\hat{S}^{JW}_j$ reduces to the identity operation. For the case when $\theta$ is $\pi$, $\hat{S}^{JW}_j$ reduces to $Z_j$. The  symmetry operation $\hat{S}$ in Jordan--Wigner encoding is now,
\begin{equation}
    \label{eq:unitary_transform_jk}
    \hat{S} = \prod_{i^*} Z_{i^*},
\end{equation}
where $i^{*}$ is the index of spin-orbitals with character -1 \cite{Yen_2019}.

\begin{figure}[ht]
\begin{center}
\includegraphics[width=\linewidth]{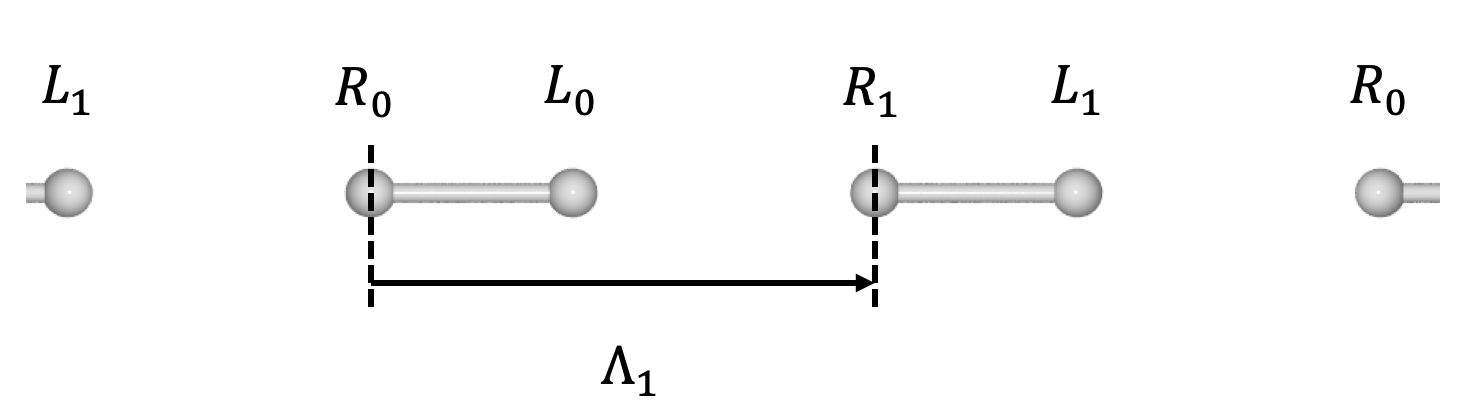}
\caption{Hydrogen molecule chain with orbital labels for $L=2$.}
\label{fig:H2k2_schematics}
\end{center}
\end{figure}

In order to find the $Z_2$ symmetry for the translation symmetry, we define the 1$s$-orbitals on a hydrogen atoms as $s^A$, where the superscript $A$ denotes the location of the orbitals as it is shown in Figure \ref{fig:H2k2_schematics}. Consistently with the transformation in Equation \ref{eq:expansion}, the Bloch orbitals are written as,
\begin{align}
    b^L_0 = s^{L_0} + s^{L_1}, \\
    b^L_1 = s^{L_0} - s^{L_1}, \\
    b^R_0 = s^{R_0} + s^{R_1}, \\
    b^R_1 = s^{R_0} - s^{R_1},
\end{align}
where the subscript of the Bloch orbitals refers to the k-points in Table \ref{tab:encodings_H2k2} and the global phase and normalisation are ignored. With the Bloch orbitals the mean-field solution results in the following crystal orbitals,
\begin{align}
    \label{eq:4qb_h2_l2_crystal_orb0}
    b^{+}_0 = b^L_0 + b^R_0,\\
    \label{eq:4qb_h2_l2_crystal_orb1}
    b^{-}_0 = b^L_0 - b^R_0,\\
    \label{eq:4qb_h2_l2_crystal_orb2}
    b^{+}_1 = b^L_1 + b^R_1,\\
    \label{eq:4qb_h2_l2_crystal_orb3}
    b^{-}_1 = b^L_1 - b^R_1,
\end{align}
where $+$ or $-$ define a bonding or anti-bonding crystal orbital. 
To find the $Z_2$ symmetry corresponding to a primitive cell translation $\Lambda_1$, we apply the translation operation $s^{L_0} \rightarrow s^{L_1}, s^{R_0} \rightarrow s^{R_1}$ to the crystal orbitals $b_0^+$, $b_0^-$, $b_1^+$ and $b_1^-$. This results in the $Z_2$ symmetry $Z_4 Z_5 Z_6 Z_7$ corresponding to the translation of the primitive cell. Similarly, if we perform the transformation $s^{L_0} \rightarrow s^{R_0}, s^{L_1} \rightarrow s^{R_1}$, we obtain the symmetry of the internal hydrogen dimer as a $Z_2$ symmetry ($Z_2 Z_3 Z_6 Z_7$). It is trivial to express alpha- and beta- number conservation as $Z_2$ symmetries. In table \ref{tab:hamiltonian_H2k2_4qb}, we state the reduced 4-qubit $L=2$ hydrogen chain Hamiltonian.

We also use the $Z_2$ symmetries defined above to reduce the number of excitations in our UCCSD-PBC ansatz. The ansatz we use for the 4-qubit $L=2$ hydrogen chain is stated in table \ref{tab:excitations_H2k2_4qb}. We enforce that all UCCSD exponents must commute with all $Z_2$ symmetries. We find the UCCSD singles do not commute, hence we remove these excitations from our ansatz. Using $Z_2$ symmetries we taper qubits on the remaining UCCSD exponents in our ansatz. In table \ref{tab:excitations_H2k2_4qb}, we state the reduced 4-qubit UCCSD-PBC ansatz.
\clearpage

\begingroup
\squeezetable
\begin{center}
\begin{longtable}{c || c c|| c c }
        \caption{8-qubit Hamiltonian $\hat{H}_K$ for Hydrogen Chain $L=2$ $d=\SI{0.75}{\angstrom}$.}
        \label{tab:hamiltonian_H2k2_8qb}

        \tabularnewline
        
        \hline
        \multicolumn{1}{c||}{}
        & \multicolumn{1}{c}{Fermion Term}
        & \multicolumn{1}{c}{Fermion Term Coefficient}
        & \multicolumn{1}{c}{Qubit Term}
        & \multicolumn{1}{c}{Qubit Term Coefficients}
        \tabularnewline
        
        \hline
        \hline
        1                                       
        & --                                     
        & $26.3123141167 + 0.0000000000\mathrm{i}$
        & --                                     
        & $2.7062283487 + 0.0000000000\mathrm{i}$
        \tabularnewline
        
        \hline
        2                                       
        & $ c^{\dagger}_{0,0,\uparrow}c_{0,0,\uparrow}$
        & $-10.9790671409 + 0.0000000000\mathrm{i}$
        & $ Z_{ 0 }$                              
        & $0.7221310735 + 0.0000000000\mathrm{i}$
        \tabularnewline
        
        \hline
        3                                       
        & $ c^{\dagger}_{0,0,\downarrow}c_{0,0,\downarrow}$
        & $-10.9790671409 + 0.0000000000\mathrm{i}$
        & $ Z_{ 1 }$                              
        & $0.7221310735 + 0.0000000000\mathrm{i}$
        \tabularnewline
        
        \hline
        4                                       
        & $ c^{\dagger}_{1,0,\uparrow}c_{1,0,\uparrow}$
        & $-10.8587001302 + 0.0000000000\mathrm{i}$
        & $ Z_{ 2 }Z_{ 3 }$                       
        & $0.7054207903 + 0.0000000000\mathrm{i}$
        \tabularnewline
\hline
        \hline
    \end{longtable}
\end{center}
\endgroup
\clearpage

\begingroup
\squeezetable
\begin{center}
    \begin{longtable}{c || c c|| c c }
        \caption{8-qubit Hamiltonian $\hat{H}_R$ for Hydrogen Chain $L=2$ $d=\SI{0.75}{\angstrom}$.}
        \label{tab:hamiltonian_H2L2_8qb}

        \tabularnewline
        
        \hline
        \multicolumn{1}{c||}{}
        & \multicolumn{1}{c}{Fermion Term}
        & \multicolumn{1}{c}{Fermion Term Coefficient}
        & \multicolumn{1}{c}{Qubit Term}
        & \multicolumn{1}{c}{Qubit Term Coefficients}
        \tabularnewline
        
        \hline
        \hline
        1                                       
        & --                                     
        & $26.3123141167 + 0.0000000000\mathrm{i}$
        & --                                     
        & $2.7062283487 + 0.0000000000\mathrm{i}$
        \tabularnewline
        
        \hline
        2                                       
        & $ a^{\dagger}_{0,0,\uparrow}a_{0,0,\uparrow}$
        & $-10.9188836355 + 0.0000000000\mathrm{i}$
        & $ Z_{ 2 }Z_{ 3 }$                       
        & $0.7174403247 + 0.0000000000\mathrm{i}$
        \tabularnewline
        
        \hline
        3                                       
        & $ a^{\dagger}_{1,0,\uparrow}a_{1,0,\uparrow}$
        & $-10.9188836355 + 0.0000000000\mathrm{i}$
        & $ Z_{ 6 }Z_{ 7 }$                       
        & $0.7174403247 + 0.0000000000\mathrm{i}$
        \tabularnewline
        
        \hline
        4                                       
        & $ a^{\dagger}_{0,0,\downarrow}a_{0,0,\downarrow}$
        & $-10.9188836355 + 0.0000000000\mathrm{i}$
        & $ Z_{ 1 }Z_{ 6 }$                       
        & $0.7090737227 + 0.0000000000\mathrm{i}$
        \tabularnewline
\hline
        \hline
    \end{longtable}
\end{center}
\endgroup
\clearpage

\begingroup
\squeezetable
\begin{center}
    \begin{longtable}{ c c c }
        \caption{4-qubit Hamiltonian $\hat{H}_K$ for Hydrogen Chain $L=2$ $d=\SI{0.75}{\angstrom}$}
        \label{tab:hamiltonian_H2k2_4qb}

        \tabularnewline
        \multicolumn{1}{c}{}
        & \multicolumn{1}{c}{Qubit Term}
        & \multicolumn{1}{c}{Qubit Term Coefficients}
        \tabularnewline
        
        \hline
        \hline
        0                                       
        & $ Z_{ 2 }Z_{ 3 }$                       
        & $2.7744459496 + 0.0000000000\mathrm{i}$
        \tabularnewline
        
        \hline
        1                                       
        & $ Z_{ 0 }Z_{ 2 }$                       
        & $2.7668087156 + 0.0000000000\mathrm{i}$
        \tabularnewline
        
        \hline
        2                                       
        & $ Z_{ 0 }Z_{ 1 }Z_{ 2 }Z_{ 3 }$         
        & $2.7660138330 + 0.0000000000\mathrm{i}$
        \tabularnewline
        
        \hline
        3                                       
        & $ Z_{ 1 }Z_{ 2 }$                       
        & $2.7595490613 + 0.0000000000\mathrm{i}$
        \tabularnewline
        
        \hline
        4                                       
        & $ Z_{ 1 }Z_{ 3 }$                       
        & $2.7271696573 + 0.0000000000\mathrm{i}$
        \tabularnewline
        
        \hline
        5                                       
        & --                                    
        & $2.7062283487 + 0.0000000000\mathrm{i}$
        \tabularnewline
\hline
        \hline

    \end{longtable}
\end{center}
\endgroup

\clearpage

\begingroup
\squeezetable
\begin{table}
    \caption{Cluster amplitudes for the hydrogen chain $d=\SI{0.75}{\angstrom}$, $L=2$ case in momentum space.}
    \label{tab:excitations_H2k2_4qb}
    \begin{ruledtabular}
        \begin{tabular}{ccccc}
        \multicolumn{1}{c}{$t_{\mathbf{k}_{q},q}^{\mathbf{k}_{p},p}$ , $t_{\mathbf{k}_{q},q,\mathbf{k}_{s},s}^{\mathbf{k}_{p},p,\mathbf{k}_{r},r}$} \footnote{Since it is a 1D system, the $\mathbf{k}$ is represented by the number $k^{(1)}$ in the amplitude symbol.}
        & \multicolumn{1}{c}{Initial}
        & \multicolumn{1}{c}{CCSD}
        & \multicolumn{1}{c}{UCCSD-PBC}
        & \multicolumn{1}{c}{Reduced 4-qubit UCCSD-PBC}
        \tabularnewline
        
        \hline
        $t_{0, 0}^{0, 1}$                   
        & $0.0000000000$
        & $0.0000000000$
        & $0.0000000000$
        & --
        \tabularnewline
        
        \hline
        $t_{1, 0}^{1, 1}$                   
        & $0.0000000000$
        & $0.0000000000$
        & $0.0000000000$
        & --
        \tabularnewline
        
        \hline
        $t_{0, 0, 0, 0}^{0, 1, 0, 1}$       
        & $-0.0161539497$
        & $-0.0222876791$
        & $-0.0229664536$
        & $-0.0229681615$
        \tabularnewline
        
        \hline
        $t_{0, 0, 0, 0}^{1, 1, 1, 1}$       
        & $-0.0569609701$
        & $-0.0823982633$
        & $-0.0816454412$
        & $-0.0816407724$
        \tabularnewline
        
        \hline
        $t_{0, 0, 1, 0}^{0, 1, 1, 1}$       
        & $0.0212597555$
        & $0.0380474459$
        & $0.0378888082$
        & $0.0376957068$\footnote{A factor $\frac{1}{2}\times$ is applied to compensate the removal of redundant excitations.}
        \tabularnewline
        
        \hline
        $t_{0, 0, 1, 0}^{1, 1, 0, 1}$       
        & $0.0395001691$
        & $0.0576759056$
        & $0.0573221895$
        & $0.0575539623$$^{\text{b}}$
        \tabularnewline
        
        \hline
        $t_{1, 0, 0, 0}^{0, 1, 1, 1}$\footnote{Redundant excitations.}       
        & $0.0395001691$
        & $0.0576759056$
        & $0.0573221895$
        & --
        \tabularnewline
\end{tabular}
    \end{ruledtabular}
\end{table}
\endgroup

\begingroup
\squeezetable
\begin{table}
    \caption{Cluster amplitudes for the hydrogen chain $d=\SI{0.75}{\angstrom}$, $L=2$ case in real space.}
    \label{tab:excitations_H2L2_4qb}
    \begin{ruledtabular}
        \begin{tabular}{cddd}
        \multicolumn{1}{c}{$t_{\mathbf{R}_{q},q}^{\mathbf{R}_{p},p}$, $t_{\mathbf{R}_{q},q,\mathbf{R}_{s},s}^{\mathbf{R}_{p},p,\mathbf{R}_{r},r}$} \footnote{Since it is a 1D system, the $\mathbf{R}$ is represented by the number $l^{(1)}$ in the amplitude symbol.}
        & \multicolumn{1}{c}{Initial}
        & \multicolumn{1}{c}{CCSD}
        & \multicolumn{1}{c}{UCCSD-PBC}
        \tabularnewline
        
        \hline
        $t_{0, 0}^{0, 1}$                       
        & 0.0000000000
        & 0.0000000000
        & 0.0000000000
        \tabularnewline
        
        \hline
        $t_{0, 0}^{1, 1}$                       
        & 0.0000000000
        & 0.0000000000
        & 0.0000000000
        \tabularnewline
        
        \hline
        $t_{1, 0}^{0, 1}$                       
        & 0.0000000000
        & 0.0000000000
        & 0.0000000000
        \tabularnewline
        
        \hline
        $t_{1, 0}^{1, 1}$                       
        & 0.0000000000
        & 0.0000000000
        & 0.0000000000
        \tabularnewline
        
        \hline
        $t_{0, 0, 0, 0}^{0, 1, 0, 1}$           
        & -0.0055512163
        & -0.0137359314
        & -0.0136913973
        \tabularnewline
        
        \hline
        $t_{0, 0, 0, 0}^{0, 1, 1, 1}$           
        & 0.0108396392
        & 0.0294891060
        & 0.0287413413
        \tabularnewline
        
        \hline
        $t_{0, 0, 0, 0}^{1, 1, 0, 1}$           
        & 0.0108396392
        & 0.0294891060
        & 0.0287413413
        \tabularnewline
\end{tabular}
    \end{ruledtabular}
\end{table}
\endgroup

\clearpage
\section{Lattice parameters and geometries}

\begin{table}[h]
\caption{Geometries of the primitive cell for the investigated systems.}
\begin{tabular}{|c|c|c|c|c|c|}
\hline 
System & Dimension & atoms & x & y & z\tabularnewline
\hline 
\hline 
\multirow{6}{*}{H$_{2}$} & \multirow{2}{*}{1D} & H1 & $0$ & $0$ & $0$\tabularnewline
\cline{3-6} \cline{4-6} \cline{5-6} \cline{6-6} 
 &  & H2 & $d$ & $0$ & $0$\tabularnewline
\cline{2-6} \cline{3-6} \cline{4-6} \cline{5-6} \cline{6-6} 
 & \multirow{2}{*}{2D} & H1 & $0$ & $0$ & $0$\tabularnewline
\cline{3-6} \cline{4-6} \cline{5-6} \cline{6-6} 
 &  & H2 & $d$ & $0$ & $0$\tabularnewline
\cline{2-6} \cline{3-6} \cline{4-6} \cline{5-6} \cline{6-6} 
 & \multirow{2}{*}{3D} & H1 & $0$ & $0$ & $0$\tabularnewline
\cline{3-6} \cline{4-6} \cline{5-6} \cline{6-6} 
 &  & H2 & $d$ & $0$ & $0$\tabularnewline
\hline 
\multirow{3}{*}{He} & 1D & He1 & $0$ & $0$ & $0$\tabularnewline
\cline{2-6} \cline{3-6} \cline{4-6} \cline{5-6} \cline{6-6} 
 & 2D & He1 & $0$ & $0$ & $0$\tabularnewline
\cline{2-6} \cline{3-6} \cline{4-6} \cline{5-6} \cline{6-6} 
 & 3D & He1 & $0$ & $0$ & $0$\tabularnewline
\hline 
\multirow{2}{*}{LiH} & \multirow{2}{*}{1D} & Li1 & $0$ & $0$ & $0$\tabularnewline
\cline{3-6} \cline{4-6} \cline{5-6} \cline{6-6} 
 &  & H1 & $d$ & $0$ & $0$\tabularnewline
\hline 
\end{tabular}

\label{tab:geometries}
\end{table}
\begin{table}[h]
\caption{Lattice vectors of the primitive cell for the investigated systems.}
\begin{tabular}{|c|c|c|c|c|c|}
\hline 
System & Dimension & Lattice vector & x & y & z\tabularnewline
\hline 
\hline 
\multirow{6}{*}{H$_{2}$} & 1D & $\mathbf{a}_{1}$ & $2.5d$ & $0$ & $0$\tabularnewline
\cline{2-6} \cline{3-6} \cline{4-6} \cline{5-6} \cline{6-6} 
 & \multirow{2}{*}{2D} & $\mathbf{a}_{1}$ & $2.5d$ & $0$ & $0$\tabularnewline
\cline{3-6} \cline{4-6} \cline{5-6} \cline{6-6} 
 &  & $\mathbf{a}_{2}$ & $0$ & $2.5d$ & $0$\tabularnewline
\cline{2-6} \cline{3-6} \cline{4-6} \cline{5-6} \cline{6-6} 
 & \multirow{3}{*}{3D} & $\mathbf{a}_{1}$ & $2.5d$ & $0$ & $0$\tabularnewline
\cline{3-6} \cline{4-6} \cline{5-6} \cline{6-6} 
 &  & $\mathbf{a}_{2}$ & $0$ & $2.5d$ & $0$\tabularnewline
\cline{3-6} \cline{4-6} \cline{5-6} \cline{6-6} 
 &  & $\mathbf{a}_{3}$ & $0$ & $0$ & $2.5d$\tabularnewline
\hline 
\multirow{6}{*}{He} & 1D & $\mathbf{a}_{1}$ & $d$ & $0$ & $0$\tabularnewline
\cline{2-6} \cline{3-6} \cline{4-6} \cline{5-6} \cline{6-6} 
 & \multirow{2}{*}{2D} & $\mathbf{a}_{1}$ & $d$ & $0$ & $0$\tabularnewline
\cline{3-6} \cline{4-6} \cline{5-6} \cline{6-6} 
 &  & $\mathbf{a}_{2}$ & $0$ & $d$ & $0$\tabularnewline
\cline{2-6} \cline{3-6} \cline{4-6} \cline{5-6} \cline{6-6} 
 & \multirow{3}{*}{3D} & $\mathbf{a}_{1}$ & $d$ & $0$ & $0$\tabularnewline
\cline{3-6} \cline{4-6} \cline{5-6} \cline{6-6} 
 &  & $\mathbf{a}_{2}$ & $0$ & $d$ & $0$\tabularnewline
\cline{3-6} \cline{4-6} \cline{5-6} \cline{6-6} 
 &  & $\mathbf{a}_{3}$ & $0$ & $0$ & $d$\tabularnewline
\hline 
LiH & 1D & $\mathbf{a}_{1}$ & $2d$ & $0$ & $0$\tabularnewline
\hline 
\end{tabular}

\label{tab:lattice_vectors}
\end{table}

\end{document}